\documentclass[usegraphicx,usenatbib]{mn2e}

\usepackage[total={17.8cm,24.0cm},centering]{geometry}
\usepackage{times}
\usepackage{amssymb}

\newcommand{\dd}{\mathrm{d}}
\newcommand{\apj}{ApJ}           % Astrophysical Journal
\newcommand{\apjl}{ApJ}           % Astrophysical Journal
\newcommand{\apjs}{ApJS}           % Astrophysical Journal
\newcommand{\mnras}{MNRAS}       % Monthly Notices of the RAS
\newcommand{\nat}{Nature}
\newcommand{\aap}{A\&A}

\newcommand{\aj}{AJ}
\newcommand{\pasj}{PASJ}

\newcommand{\sauron}{\texttt{SAURON}}

\newcommand{\re}{\hbox{$R_{\rm e}$}}

\newcommand{\vse}{\hbox{$(V/\sigma,\varepsilon)$}}
\newcommand{\plotone}[1]{\includegraphics[width=\columnwidth]{#1}}
\newcommand{\refsec}[1]{Section~\ref{#1}}
\newcommand{\reffig}[1]{Fig.~\ref{#1}}
\newcommand{\refeq}[1]{equation~(\ref{#1})}
\newcommand{\text}[1]{\textrm{#1}}

\title[Anisotropic axisymmetric Jeans models]
{Measuring the inclination and mass-to-light ratio of axisymmetric galaxies via anisotropic Jeans models of stellar kinematics}

\author[M.~Cappellari]{Michele Cappellari\thanks{E-mail:
cappellari@astro.ox.ac.uk}\\
Sub-Department of Astrophysics, University of Oxford, Denys Wilkinson Building, Keble Road, Oxford, OX1 3RH}

\date{Accepted 2008 July 23. Received 2008 July 17; in original form 2008 May 30}

\volume{390}
\pagerange{71--86}
\pubyear{2008}

\begin{document}
\maketitle

\begin{abstract}
We present a simple and efficient anisotropic generalization of the semi-isotropic (two-integral) axisymmetric Jeans formalism which is used to model the stellar kinematics of galaxies. The following is assumed: (i) a constant mass-to-light ratio $M/L$ and (ii) a velocity ellipsoid that is aligned with cylindrical coordinates $(R,z)$ and characterized by the classic anisotropy parameter $\beta_z=1-\overline{v_{z}^2}/\overline{v_{R}^2}$. Our simple models are fit to \sauron\ integral-field observations of the stellar kinematics for a set of fast-rotator early-type galaxies.
With only two free parameters ($\beta_z$ and the inclination) the models generally provide remarkably good descriptions of the shape of the first ($V$) and second ($V_{\rm rms}\equiv\sqrt{V^2+\sigma^2}$) velocity moments, once a detailed description of the surface brightness is given.
This is consistent with previous findings on the dynamical structure of these objects. With the observationally-motivated assumption that $\beta_z\ga0$, the method is able to recover the inclination. The technique can be used to determine the dynamical mass-to-light ratios and angular momenta of early-type fast-rotators and spiral galaxies, especially when the quality of the data does not justify more sophisticated modeling approaches. This formalism allows for the inclusion of dark matter, supermassive black holes, spatially varying anisotropy, and multiple kinematic components.
\end{abstract}

\begin{keywords}
galaxies: elliptical and lenticular, cD --
galaxies: evolution -- galaxies: formation -- galaxies: kinematics and
dynamics -- galaxies: structure
\end{keywords}

\section{Introduction}

According to the theory that best reproduces the observations, galaxy formation proceeds in a hierarchical fashion, driven by gravity, in a Universe dominated by dark matter of unknown nature \citep[e.g.][]{springel05nat}. The hierarchy of merging is accompanied by changes in galaxy structure and morphology. In particular early-type galaxies (Es and S0s) are thought to form by the gas-rich merging of spiral galaxies or by gas starvation of spirals, followed by subsequent collisionless mergers \citep[e.g.][]{faber07}.

Three key global parameters can be used to characterize galaxies structure while studying this sequence of merging of galaxies and dark matter: (i) The angular momentum, which varies during mergers and increases with the amount of gas dissipation, (ii) the stellar population, which records the history of star formation events during the gas-rich mergers, and (iii) the mass-to-light ratio, which is affected by both the population and by the dark-matter fraction.

The large majority of the galaxies in the Universe are to first order axisymmetric (except for bars) and posses disks. This includes fast-rotator early-type galaxies \citep{emsellem07,cappellari07} and spiral galaxies. Both the fast-rotator early-types \citep{gerhard01,rusin05,cappellari06,koopmans06,thomas07} and the spiral galaxies \citep{persic96,palunas00,bell01,kassin06} appear dominated by the stellar matter inside one half-light radius. Observations suggest that they have a dynamical structure characterized by a flattening of the velocity ellipsoid in the $z$ direction parallel to the galaxy symmetry axis \citep{gerssen97,gerssen00,shapiro03,cappellari07,Noordermeer08}.

The goal of this paper is to include the knowledge of the structure of the fast-rotator and spiral galaxies, into a realistic but simple dynamical modeling method, which can be applied to the measurement of both the mass-to-light ratio and the amount of rotation (for which the inclination is needed) in the central regions of these galaxies, while also allowing for the inclusion of dark matter and the study of multiple kinematical components. The success of the adopted model's assumptions in describing the kinematics of real galaxies is verified against integral-field observations of the stellar kinematics obtained with \sauron\ \citep{bacon01}.

In \refsec{sec:solving_jeans} we briefly review the theory and past applications of the Jeans equations and of the shape of the velocity ellipsoid in galaxies. In \refsec{sec:anisotropic_jeans_solutions} we describe our new anisotropic Jeans formalism, which we apply and test in \refsec{sec:tests}. Finally our results are summarized in \refsec{sec:summary}.

\section{Solving the axisymmetric Jeans equations}
\label{sec:solving_jeans}

\subsection{The collisionless Boltzmann equation}

The positions $\mathbf{x}$ and velocities $\mathbf{v}$ of a large system of stars can be described by the distribution function (DF) $f(\mathbf{x},\mathbf{v})$. When the system is in a {\em steady state} under the gravitational influence of a smooth potential $\Phi$, the DF must satisfy the fundamental equation of stellar dynamics, the steady-state collisionless Boltzmann equation (\citealt{binney87}, hereafter BT; equation [4-13b])
\begin{equation}
\sum _{i=1}^3 \left(v_i
   \frac{\partial f}{\partial
   x_i}-\frac{\partial \Phi
   }{\partial x_i}
   \frac{\partial f}{\partial
   v_i}\right)=0.
   \label{eq:boltzmann}
\end{equation}
Given that $f$ is a function of six variables, an infinite family of solutions satisfies \refeq{eq:boltzmann}. Additional assumptions and simplifications are required for a practical application of the equation. One classic way of constraining the problem consists of drastically reducing it, from that of recovering the DF, to that of studying only the velocity moments of the DF. This approach leads to the Jeans equations, which are discussed in the next section.

\subsection{The Jeans equations}

\subsubsection{Summary of derivation}

If we rewrite \refeq{eq:boltzmann} in standard cylindrical coordinates  $(R,z,\phi)$ and we make the important assumption of {\em axial symmetry} ($\partial \Phi/\partial\phi=\partial f/\partial\phi=0$) we obtain (cf.\ BT equation~[4-17])
\begin{equation}
    v_R \frac{\partial f}{\partial R}
    + v_z \frac{\partial f}{\partial z}
    + \left(\frac{v_\phi^2}{R} - \frac{\partial \Phi}{\partial R}\right) \frac{\partial f}{\partial v_R}
    - \frac{\partial \Phi }{\partial z} \frac{\partial f}{\partial v_z}
    -\frac{v_R v_\phi}{R} \frac{\partial f}{\partial v_{\phi}}
    = 0.
    \label{eq:boltz_cyl}
\end{equation}
Multiplying \refeq{eq:boltz_cyl} respectively by $v_R$ and by $v_z$, and integrating over all velocities, we obtain the two\footnote{All terms in the third Jeans equation, involving a multiplication by $v_\phi$, are zero in axisymmetry.} Jeans equations (\citealt{jeans22}; BT equation [4-29a,c])
\begin{eqnarray}
    \frac{\nu\overline{v_R^2}-\nu\overline{v_\phi^2}}{R}
    + \frac{\partial(\nu\overline{v_R^2})}{\partial R}
    + \frac{\partial(\nu\overline{v_R v_z})}{\partial z}
    & = & -\nu\frac{\partial\Phi}{\partial R}
    \label{eq:jeans_cyl_R}\\
    \frac{\nu\overline{v_R v_z}}{R}
    + \frac{\partial(\nu\overline{v_z^2})}{\partial z}
    + \frac{\partial(\nu\overline{v_R v_z})}{\partial R}
    & = & -\nu\frac{\partial\Phi}{\partial z}.
    \label{eq:jeans_cyl_z}
\end{eqnarray}
where we use the notation
\begin{equation}
    \nu\overline{v_k v_j}\equiv\int v_k v_j f\; \dd^3 \mathbf{v}.
\end{equation}

These equations are still quite general, as they derive from the steady-state Boltzmann \refeq{eq:boltzmann} with the only assumption of axisymmetry. They do {\em not} require self consistency (a potential $\Phi$ generated by the luminosity density $\nu$) and they make no assumption on the DF. However, even if one assumes $\Phi$ to be known (it may be derived from the observed $\nu$ via the Poisson equation), the two equations~(\ref{eq:jeans_cyl_R}) and (\ref{eq:jeans_cyl_z}) are still a function of the four unknown $\overline{v_R^2}$, $\overline{v_z^2}$, $\overline{v_\phi^2}$ and $\overline{v_R v_z}$ and do not uniquely specify a solution.

\subsubsection{Closing the axisymmetric Jeans equations}

Given that the axisymmetric Jeans equations relate four functions of the meridional plane $(R,z)$ coordinates, one needs to specify at least two functions of $(R,z)$ for a unique solution. A natural way to provide a unique solution for the Jeans equations consists of specifying the shape and orientation of the intersection of the velocity ellipsoid everywhere in the meridional plane. In fact the three components $\overline{v_R^2}$, $\overline{v_z^2}$ and $\overline{v_R v_z}$, of the velocity dispersion tensor, uniquely describe the equation of the velocity dispersion ellipse, whose shape can be derived by diagonalizing the tensor. The velocity dispersion ellipse will have the major axis at an angle $\theta$ with respect to the equatorial plane
\begin{equation}\label{eq:pa}
    \tan 2\theta = \frac{2\; \overline{v_R v_z}}{\overline{v_R^2}-\overline{v_z^2}}
\end{equation}
and an axial ratio $0\le q\le 1$ given by
\begin{equation}\label{eq:q}
    q^2 =
    \frac{\overline{v_R^2}+\overline{v_z^2} - \sqrt{(\overline{v_R^2}-\overline{v_z^2})^2 + 4\;\overline{v_R v_z}^2}}
    {\overline{v_R^2}+\overline{v_z^2} + \sqrt{(\overline{v_R^2}-\overline{v_z^2})^2 + 4\;\overline{v_R v_z}^2}}.
\end{equation}
The specification of the ellipse geometry ($\theta$ and $q$) is then sufficient to write two of the three variables as a function of the remaining one.

The most common choice which is made consists of assuming\footnote{In the original paper by \citet{jeans22} this was {\em not} an assumption. Stellar orbits were thought to conserve only two isolating integrals of motion, in which case the DF naturally possess that special semi-isotropic form.} a circular velocity-ellipsoid in the meridional plane (sometimes called semi-isotropy condition). This assumption implies $\overline{v_R^2}=\overline{v_z^2}$ and $\overline{v_R v_z}=0$, and is sufficient to `close' the set of equations to provide a unique solution for the remaining variables $\overline{v_z^2}$ and $\overline{v_\phi^2}$. Other common options are discussed in \refsec{sec:choice}.

A well studied special case of semi-isotropic system is one in which the 3-integral DF which generally characterizes a stationary system, is assumed to depend only on the two classical integrals of motion $f=f(E,L_z)$, where $E$ is the potential energy, and $L_z$ is the angular momentum with respect to the symmetry $z$-axis. For this reason the semi-isotropic Jeans models are often called two-integral Jeans models.

In this paper we are investigating whether it is possible to make an alternative assumption on the shape of the velocity ellipsoid, which retains the simplicity of the widely used semi-isotropic assumption, while also providing a better description of real galaxies and still leading to an efficient solution of the Jeans equations.

\subsubsection{Past applications of the semi-isotropic Jeans equations}
\label{sec:review}

Due to its simplicity, the semi-isotropy assumption has proven remarkably useful in a large variety of applications of the Jeans equations. It was used to quantify the amount of rotation in galaxies \citep{nagai76,satoh80,binney90,vanDerMarel90}, to search for hidden disks in elliptical galaxies \citep{rix92,cinzano94,rix99}, to measure their mass-to-light ratio \citep{vanDerMarel91,statler99,cappellari06,cortes08} and dark matter profiles \citep{carollo95}, to study the connection between line-strength in galaxies and the local escape velocities \citep{davies93,carollo94}, to study the scatter in the Fundamental Plane \citep{vanAlbada95,lanzoni03,riciputi05}, to measure the deviations of the gas kinematics from circular velocities \citep{bertola95,corsini99,cinzano99,young08}, and to estimate the masses of supermassive black holes \citep{magorrian98,vanDerMarel98,cretton99ngc4342,joseph01}. Although these models have been largely superseded by \citet{schwarzschild79} orbit-superposition method, when good kinematic data are available and the maximum generality is required, they are becoming useful to study the mass-to-light ratios and rotation of galaxies at high-redshift, where the data quality still does not justify more sophisticated approaches \citep{vanDerMarel07a,vanDerMarel07b,vanderWel08}.

\subsection{Shape and orientation of the velocity ellipsoid in galaxies}

To explore the possibility of making a simple but sufficiently realistic assumption on the shape and orientation of the velocity ellipsoid in the meridional plane of axisymmetric galaxies, we need to understand what that shape is expected to be.

\subsubsection{Theory}

A qualitative insight into the orientation of the velocity ellipse in real galaxies can start from the analysis of the special case of separable potentials \citep{dezeeuw85}. In an axisymmetric oblate separable potential the equations of motion for the stellar obits can be separated in a prolate spheroidal coordinates system $(\lambda,\phi,\nu)$ which also defines the boundaries of the orbits. In other words the orbital motions can be written as the linear combination of two independent oscillations in $\lambda$ and $\nu$, plus a non-uniform rotation around the symmetry axis. For this reason the velocity ellipse is aligned with the coordinate system  and the cross term $\overline{v_\lambda v_\nu}$ vanishes \citep{eddington15,dezeeuw90}. The prolate spheroidal coordinates are characterized by the fact that they tend to be aligned with the cylindrical coordinates $(R,z,\phi)$ at small radii and with the polar ones $(r,\theta,\phi)$ at large radii. Separable potentials are characterized by a central constant-density region. The cylindrical alignment happens in that region, as the stars there tend to move like an harmonic oscillator. For these reasons one expects the velocity dispersion ellipse to be cylindrically aligned at small radii and radially aligned at large radii.

The gravitational potential of real galaxies is {\em not} of the separable form. Separable potentials are in fact necessarily characterized by smooth analytic centers, while real galaxies possess central singularities due to cusped density profiles \citep[e.g.][]{ferrarese94,lauer95} and supermassive black holes \citep{magorrian98,gebhardt00bhs,ferrarese00}. Still, numerical integrations of orbits in non-separable axisymmetric potentials show that most of them are still bounded by curves, which qualitatively resemble the spheroidal coordinates \citep[e.g.][]{richstone82,dehnen93,cretton99}. For this reason one can expect the velocity dispersion ellipse to behave in real galaxies in a way that is qualitatively similar to the separable case. Numerical calculations confirm this fact in non-separable triaxial and axisymmetric potentials \citep{merritt80,dehnen93}. In the limit of a point mass, the velocity ellipsoid has to be spherically aligned for symmetry. This suggests that at small radii, where the supermassive black hole or the stellar cusp dominates, the velocity ellipsoid in real galaxies should be more spherically oriented than in the separable case.

\subsubsection{Observations}

\begin{figure}
  \plotone{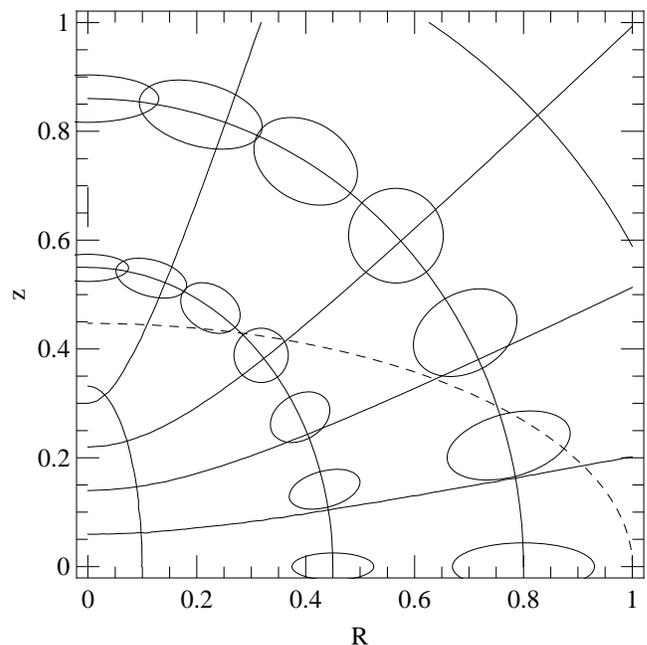}
  \caption{Qualitative description of the shape and orientation of the velocity ellipsoid in the meridional plane $(R,z)$ of the fast-rotator early-type galaxies (derived from \citealt{cappellari07}). The ellipses show the intersection of the velocity ellipsoid with the $(v_R,v_z)$ plane. The solid lines show a representative prolate spheroidal system of coordinates. The dashed line is a representative iso-density contour. The ellipsoids are aligned in spheroidal coordinates, but the axial ratio of the ellipses varies with the polar angle in such a way that the shape and orientation of the ellipsoids are similar on both the equatorial plane and the symmetry axis.}\label{fig:coordinates}
\end{figure}

The DF of a stationary system is a function of the three separable integrals of motion \citep{jeans15}. In general it cannot be recovered from real galaxies without at least another three-dimensional observable quantity. This quantity is now being provided by integral-field observations of the stellar kinematics \citep[e.g.][]{emsellem04}, which allow the stellar line-of-sight velocity-distribution to be measured at every position of the galaxy image on the sky. We used these observations, in combination with orbit-based three-integral axisymmetric models, to measure the shape and orientation of the velocity ellipsoid at every position within the meridional plane (within one half-light radius \re), for a sample of 25 early-type galaxies \citep{cappellari07}. \reffig{fig:coordinates} qualitatively summarizes the main findings of that paper regarding the shape of the velocity dispersion ellipsoid for the fast-rotator galaxies: (i) The ellipsoids qualitatively behave like in separable potentials and already near 1\re\ they are essentially spherically oriented; (ii) the axial ratio of the ellipsoids varies gradually as a function of the polar angle, in such a way that the ellipsoid has nearly the same shape on both the equatorial plane (where the density is the highest) and the symmetry axis. The net effect is that to first order the global anisotropy of the galaxies can be described as a simple flattening of the velocity ellipsoid in the $z$-direction ($\overline{v_z^2}<\overline{v_R^2}$).

\subsection{Choice of the coordinate system for the Jeans solution}
\label{sec:choice}

One can think of three physically motivated choices of the coordinates system in which to align the velocity ellipsoid, for a unique solution of the axisymmetric Jeans equations. In this section we examine advantages and problems of each one in turn.

\subsubsection{Prolate spheroidal coordinates}
\label{sec:spheroidal}

The observed behavior of the ellipsoid could be approximated by solving the Jeans equations in prolate spheroidal coordinates in a generic axisymmetric potential \citep{dejonghe88,evans91,arnold95,vanDeVen03}. To qualitatively reproduce the behavior of \reffig{fig:coordinates}, the simplest realistic model would need at least four free parameters. Two parameters are required to define the coordinate system. A third parameter could define the shape of the velocity ellipse on the equatorial plane and another parameter would describe the angular variation. A problem of this approach is that the general solution, to derive the observables for realistic galaxy potentials, requires at least a computationally-expensive triple\footnote{For typical accuracies, every additional numerical quadrature generally increases the computation time by about {\em two} orders of magnitude.} numerical quadrature. The derived equations are also rather cumbersome and not justified by the fact that these models can at best provide a qualitative description of real galaxies. For these reasons those solutions have currently only been applied to a handful of analytic potentials, and no application to real galaxies exist.

\subsubsection{Spherical coordinates}

Given that the velocity ellipsoids in \reffig{fig:coordinates} are nearly spherically aligned, a much simpler choice would consist of solving the anisotropic Jeans equation in spherical coordinates. A solution for this case was presented by \citet{bacon83} and applied to real galaxies by \citet{bacon85} and \citet{fillmore86}. However even under this spherical assumption the solution still has the same form as in the spheroidal coordinates. The computation still involves a triple numerical quadrature. This is in fact a generic feature of the axisymmetric Jeans solution: even in the simple case in which the density is assumed to be stratified on similar oblate spheroids, one quadrature is needed to obtain the potential, a second quadrature provides the intrinsic velocity's second moments and a third quadrature finally gives the projected observables. In the spherically-oriented case the coordinates system has no free parameters and the simplest realistic model to reproduce \reffig{fig:coordinates} could have two parameters (anisotropy and its angular variation).

\subsubsection{Cylindrical coordinates}

A more radical alternative is to assume the velocity ellipse is oriented with the cylindrical coordinate system. This option was tested by e.g.\ \citet{fillmore86}, however until now, for the reasons described in  \refsec{sec:spheroidal}, it was not considered a sensible choice to describe the shape of the velocity ellipsoid in real galaxies. Our new results, based on the \sauron\ integral-field observations \citep{cappellari07}, make this choice worth exploring again. This option cannot provide a formally accurate representation of \reffig{fig:coordinates}, however it is accurate near the equatorial plane, where the density is at its maximum, and near the minor axis. Models with cylindrically-oriented velocity ellipsoid provide a good qualitative description of the empirical observation that the global anisotropy in fast-rotator galaxies is best characterized as a flattening of the velocity ellipsoid in the vertical $z$-direction \citep[their Fig.~2]{cappellari07}. The near-cylindrical orientation of the velocity ellipsoid may be due to the presence of disks, where this orientation appears natural. For this reason the cylindrical orientation is certainly appropriate to describe the dynamics of spiral galaxies \citep{gerssen97,gerssen00,shapiro03,Noordermeer08}.

In the next section we show that the assumption of a cylindrically-aligned velocity-dispersion ellipsoid, combined with the powerful Multi-Gaussian Expansion (MGE) method of \citet{emsellem94}, can generate simple solutions that well reproduce the integral-field kinematics of real galaxies, and also allow for variable mass-to-light ratios (e.g. dark matter), spatially varying anisotropy and multiple kinematical components. All this while still requiring a {\em single} numerical quadrature to predict the observables on the sky plane (or any other projection).

\section{Anisotropic Jeans solutions}
\label{sec:anisotropic_jeans_solutions}

\subsection{Axisymmetric case}
\label{sec:cylindrical_jeans}

\subsubsection{General solution}
\label{sec:jeans_solution}

We start from the general axisymmetric Jeans equations (\ref{eq:jeans_cyl_R}) and (\ref{eq:jeans_cyl_z}) and we make the following two assumptions: (i) the velocity ellipsoid is aligned with the cylindrical coordinate system $(R,z,\phi)$ and (ii) the anisotropy is constant and quantified by $\overline{v_R^2}=b\,\overline{v_z^2}$. In this case the Jeans equations reduce to
\begin{eqnarray}
    \frac{b\,\nu\overline{v_z^2}-\nu\overline{v_\phi^2}}{R}
    + \frac{\partial(b\,\nu\overline{v_z^2})}{\partial R}
    & = & -\nu\frac{\partial\Phi}{\partial R}
    \label{eq:jeans_beta_R}\\
    \frac{\partial(\nu\overline{v_z^2})}{\partial z}
    & = & -\nu\frac{\partial\Phi}{\partial z},
    \label{eq:jeans_beta_z}
\end{eqnarray}
which corresponds to the semi-isotropic case (two-integral) when $b=1$. With the boundary condition $\nu\overline{v_z^2}=0$ as $z\rightarrow\infty$
the solution reads
\begin{eqnarray}
    \nu\overline{v_z^2}(R,z)
    & = & \int_z^\infty \nu\frac{\partial\Phi}{\partial z}\dd z
    \label{eq:jeans_sol_z}\\
    \nu\overline{v_\phi^2}(R,z) & = &
    b\left[
    R \frac{\partial(\nu\overline{v_z^2})}{\partial R}
    + \nu\overline{v_z^2} \right]
    + R \nu\frac{\partial\Phi}{\partial R}
    \label{eq:jeans_sol_R}.
\end{eqnarray}
A general caveat regarding the Jeans equations is that the existence of a solution does not guarantee the existence of a corresponding physical positive DF. This can only be verified using different techniques.

\subsubsection{Summary of MGE formalism}
\label{sec:mge_formalism}

To derive solutions for the Jeans equations we make an explicit choice for the parametrization of the stellar density and the total density (which can include dark matter and a central black hole). We adopt for {\em both} the MGE parametrization of \citet{emsellem94} due to its flexibility in accurately reproducing the surface-brightness of real galaxies and the availability of robust routines\footnote{Available from http://www-astro.physics.ox.ac.uk/$\sim$mxc/idl/} to fit the galaxy photometry \citep{cappellari02mge}. If the $x$-axis is aligned with the photometric major axis, the surface brightness $\Sigma$ at the location $(x',y')$ on the plane of the sky can be written as
\begin{equation}\label{eq:surf}
\Sigma(x',y') = \sum_{k=1}^N
{\frac{L_k}{2\pi\sigma^2_k q_k'} \exp
\left[
    -\frac{1}{2\sigma^2_k}
    \left(x'^2 + \frac{y'^2}{q'^2_k} \right)
\right]},
\end{equation}
where $N$ is the number of the adopted Gaussian components, having total luminosity $L_k$, observed axial ratio $0\le q'_k\le1$ and dispersion $\sigma_k$ along the major axis.

The deprojection of the surface brightness to obtain the intrinsic luminosity density is not unique unless the axisymmetric galaxy is seen edge-on ($i=90^\circ$) \citep{rybicki87,kochanek96}, and the degeneracy becomes especially severe when the galaxy is seen at low inclinations  \citep{gerhard96,romanowsky97,vanDenBosch97,magorrian99}.
The MGE method provides a simple possible choice for the deprojection \citep{monnet92}. One of the advantages of the MGE method is that one can easily enforce the roundness of the model \citep{cappellari02mge}, thus producing realistic densities, which look like real galaxies when projected at any angle. The method cannot eliminate the intrinsic degeneracy of the deprojection and this has to be considered when interpreting results of galaxies that are close to face-on. The deprojected MGE oblate axisymmetric luminous density $\nu$ can be written as
\begin{equation}\label{eq:dens}
\nu(R,z) = \sum_{k=1}^N
\frac{L_k}{(\sqrt{2\pi}\, \sigma_k)^3 q_k} \exp
\left[
    -\frac{1}{2\sigma_k^2}
    \left(R^2+\frac{z^2}{q_k^2} \right)
\right],
\end{equation}
where the individual components have the same luminosity $L_k$ and dispersion $\sigma_k$ as in the projected case (\ref{eq:surf}), and the intrinsic axial ratio of each Gaussian becomes
\begin{equation}\label{eq:qmge}
    q_k=\frac{\sqrt{q'^2_k-\cos^2 i}}{\sin i},
\end{equation}
where $i$ is the galaxy inclination ($i=90^\circ$ being edge-on).

The total density $\rho$ can be generally described by a different set of M Gaussian components
\begin{equation}\label{eq:mass}
\rho(R,z) = \sum_{j=1}^M
\frac{M_j}{(\sqrt{2\pi}\, \sigma_j)^3 q_j} \exp
\left[
    -\frac{1}{2\sigma_j^2}
    \left(R^2+\frac{z^2}{q_j^2} \right)
\right].
\end{equation}
In the self-consistent case the Gaussians are the same as in \refeq{eq:dens} and one has $M=N$, $\sigma_j=\sigma_k$, $q_j=q_k$ and $M_j=\Upsilon_k L_k$, where $\Upsilon_k$ is the mass-to-light ratio, which can be different for different components. In the non-self-consistent case the density can be described with the sum of two sets of Gaussians: the first derived by deprojecting the surface brightness with \refeq{eq:dens}, and the second e.g. obtained by fitting a (one-dimensional) MGE model to some adopted analytic parametrization for the dark matter \citep[e.g.][]{navarro96}.

The gravitational potential generated by the density of \refeq{eq:mass} is given by \citep{emsellem94}
\begin{equation}
\Phi(R,z) = -\sqrt{2/\pi}\, G \int_0^1 \sum_{j=1}^M
{\frac{M_j\, {\mathcal H}_j(u)}{\sigma_j}} \dd u,
\label{eq:mge_potential}
\end{equation}
where $G$ is the gravitational constant and with
\begin{equation}
\mathcal{H}_j(u) = \frac{{\exp \left\{ - \frac{{u^2 }}
{{2\sigma _j^2 }}\left[ {R^2  + \frac{{z^2 }}{{1 - (1 - q_j^2 )u^2 }}}
\right] \right\}}}{{\sqrt {1 - (1 - q_j^2 )u^2 } }}.
\label{eq:integral}
\end{equation}
A supermassive black hole can be modeled by adding a Keplerian potential
\begin{equation}\label{eq:kepler}
\Phi_\bullet(R,z) = -\frac{G M_\bullet}{\sqrt{R^2+z^2}}
\end{equation}
to \refeq{eq:mge_potential}. However, as pointed out by \citet{emsellem94}, an even simpler approach consists of modeling the black hole as a small Gaussian in \refeq{eq:mge_potential}, having mass $M_j=M_\bullet$, $q_j=1$ and $3\sigma_j\lesssim r_{\rm min}$, where $r_{\rm min}$ is the smallest distance from the black hole that one needs to accurately model (e.g.\ one could choose $r_{\rm min}\approx\sigma_{\rm psf}$).

\subsubsection{MGE Jeans solution}
\label{sec:mge_solution}

Now we apply the MGE formalism to the solution of the axisymmetric anisotropic Jeans equations of \refsec{sec:jeans_solution}. Our derivation is an extension of what was done in the semi-isotropic self-consistent case ($b_k=1$ and $M_j=\Upsilon L_k$) in section~3.4 of \citet{emsellem94}. As already pointed out by \citet{jeans22}, his equations can be used to model the kinematics of different dynamical tracers, as long as they all move in the same potential \citep[e.g.][]{rix92,cinzano94}. To maintain generality we will then write the solution for the individual $N$ luminous Gaussian components, which can then be assumed to have different anisotropy. This fact can be used e.g.\ to model anisotropy gradients, or to study the anisotropy of kinematical subcomponents in galaxies (e.g.\ bulge and disk).  Substituting equations~(\ref{eq:dens}) and (\ref{eq:mge_potential}) into equations~(\ref{eq:jeans_sol_z}) and (\ref{eq:jeans_sol_R}), the integral in $z$ can be performed analytically and we obtain
\begin{eqnarray}
    \label{eq:sigma_R}
    \lefteqn{[\nu\overline{v_R^2}]_k = b_k[\nu\overline{v_z^2}]_k}\\
    \label{eq:sigma_z}
    \lefteqn{[\nu\overline{v_z^2}]_k = 4\pi G\!\! \int_0^1\!\! \sum_{j=1}^M
    \frac{\sigma_k^2 q_k^2 \nu_k q_j \rho_{0j} \mathcal{H}_j(u)\, u^2}
    {1-\mathcal{C} u^2}\, \dd u}\\
    \label{eq:vel2_phi}
    \lefteqn{[\nu\overline{v_{\phi}^2}]_k =b_k[\nu\overline{v_z^2}]_k } \nonumber\\
    && {}+ 4\pi G \int_0^1 \sum_{j=1}^M
    \frac{\nu_k q_j \rho_{0j} \mathcal{H}_j(u)\, u^2}
    {1-\mathcal{C} u^2}\mathcal{D}\,R^2 \dd u\\
    && {}=4\pi G \int_0^1 \sum_{j=1}^M
    \frac{\nu_k q_j \rho_{0j} \mathcal{H}_j(u)\, u^2}
    {1-\mathcal{C} u^2}\left(\mathcal{D}\,R^2 + b_k \sigma_k^2 q_k^2\right) \dd u\nonumber,
\end{eqnarray}
where we defined $\nu_k=\nu_k(R,z)$, $\rho_{0j}=\rho_j(0,0)$ and
\begin{equation}
    \mathcal{C}=1-q_j^2-\frac{\sigma_k^2\, q_k^2}{\sigma_j^2}
\end{equation}
\begin{equation}
    \mathcal{D}=1-b_k\, q_k^2 - \left[(1-b_k)\,\mathcal{C} + (1-q_j^2)\, b_k\right] u^2.
\end{equation}
In all the equations of this paper the index $k$ refers to the parameters, or the anisotropy, of the Gaussians describing the galaxy's luminosity density (equation~[\ref{eq:dens}]), while the index $j$ refers to the parameters of the Gaussians describing the total mass (equation~[\ref{eq:mass}]), from which the potential is obtained.

When $b_k$ is not the same for the individual luminous Gaussians, the total luminosity-weighted anisotropy at a certain spatial location $(R,z)$ of an MGE model is given by the standard definition \citep{binney82}, combined with \refeq{eq:sigma_R}:
\begin{equation}\label{eq:var_beta}
\beta_z(R,z)\equiv
1 - \frac{\overline{v_{z}^2}} {\overline{v_{R}^2}} =
1 - \frac{\sum_k [\nu\overline{v_z^2}]_k} {\sum_k b_k [\nu\overline{v_z^2}]_k}\approx
1 - \frac{\sum_k\nu_k} {\sum_k b_k\nu_k}.
\end{equation}
The last approximation comes from the fact that $[\overline{v_z^2}]_k$, being mostly a function of the total MGE potential, varies relatively little for the different Gaussians, while $\nu_k$ can be completely different and varies by many orders of magnitude for the various luminous MGE components. This allows the global anisotropy of an MGE model, at a certain spatial location in the meridional plane, to be approximately estimated from a simple luminosity-weighted sum of $b_k$. A similar reasoning applies to the estimation of the total $\kappa$ parameter in in \refsec{sec:satoh} and the total $\beta$ parameter in \refsec{sec:mge_spherical}.

\subsubsection{Line-of-sight integration of the velocity second moment}
\label{sec:los_mu2}

The intrinsic quantities have to be integrated along the line-of-sight (LOS) to generate the observables which can be compared with the galaxy kinematics. For this we define a system of sky coordinates with the $z'$ axis along the LOS and the $x'$ axis aligned with the galaxy projected major axis (see equation~[\ref{eq:surf}]). The galaxy coordinates $(x,y,z)$ are related to the ones in the sky plane by
\begin{equation}
\left( \begin{array}{l}
 x \\
 y \\
 z \\
 \end{array} \right) = \left( {\begin{array}{*{20}c}
 1 &  0           &  0         \\
 0 &  { - \cos i} &  {\sin i}  \\
 0 &  {\sin i}    &  {\cos i}  \\
\end{array}} \right)\left( \begin{array}{l}
 x' \\
 y' \\
 z' \\
 \end{array} \right),
\end{equation}
where the $z$-axis coincides with the galaxy symmetry axis and the cylindrical radius is defied by $R^2=x^2+y^2$. The projected second velocity moment along the line-of-sight $\overline{v_{\rm los}^2}\equiv\overline{v_{z'}^2}$, for one luminous Gaussian component, is then given by\footnote{Completely analogue expressions can be found for the proper motions, using e.g.\ the formulas in appendix~A of \citet{evans94}. We found that all three components of the projected proper motion dispersion tensor can be written via single quadratures and using no special functions.}
\begin{eqnarray}\label{eq:second_moments_projection}
    \lefteqn{[\Sigma\,\overline{v_{\rm los}^2}]_k = \int_{-\infty}^{\infty}
    \left\{
    [\nu\overline{v_z^2}]_k \cos^2 i \right.} \nonumber\\
    & & + \left.\left([\nu\overline{v_R^2}]_k\sin^2\phi +
    [\nu\overline{v_\phi^2}]_k\cos^2\phi\right)\sin^2 i
    \right\} \dd z',
\end{eqnarray}
where $\cos\phi=x/R$, while the total observed second moment, for the whole MGE model, will be
\begin{equation}\label{eq:sum_vz2}
    \Sigma\,\overline{v_{\rm los}^2} = \sum_{k=1}^N [\Sigma\,\overline{v_{\rm los}^2}]_k.
\end{equation}
After substitution of equations~(\ref{eq:sigma_R})--(\ref{eq:vel2_phi}), the $z'$ integral can be written explicitly. Summing over all the $N$ luminous Gaussian components we obtain the final expression
\begin{eqnarray}\label{eq:second_moment}
    \lefteqn{\Sigma\,\overline{v_{\rm los}^2}(x',y') = 4\pi^{3/2} G \int_0^1 \sum_{k=1}^N \sum_{j=1}^M\, \nu_{0k}\, q_j\, \rho_{0j}\, u^2} \nonumber \\
  & & \times\frac{
  \sigma_k^2 q_k^2 \left(\cos^2 i + b_k \sin^2 i\right) + \mathcal{D}\, x'^2\sin^2 i
  }{
  {\left(1-\mathcal{C} u^2\right) \sqrt{\left(\mathcal{A} + \mathcal{B}\cos^2 i\right)
       \left[1-(1-q_j^2)u^2\right]}}
  } \nonumber \\
  & & \times\exp\left\{-\mathcal{A}\left[x'^2+\frac{(\mathcal{A}+\mathcal{B}) y'^2}{\mathcal{A} + \mathcal{B}\cos^2 i}\right]\right\}\dd u,
\end{eqnarray}
where we defined $\nu_{0k}=\nu_k(0,0)$ and
\begin{equation}\label{eq:a}
    \mathcal{A}=\frac{1}{2}\left(\frac{u^2}{\sigma_j^2} + \frac{1}{\sigma_k^2}\right)
\end{equation}
\begin{equation}\label{eq:b}
    \mathcal{B}=\frac{1}{2}\left\{\frac{1-q_k^2}{\sigma_k^2 q_k^2}
    +\frac{(1-q_j^2)u^4}{\sigma_j^2\left[1-(1-q_j^2)u^2\right]}\right\}
\end{equation}
As expected \refeq{eq:second_moment} reduces\footnote{As in \citet{cappellari06} we corrected the typo in the expression for $B$ in equation~(63) of \citet{emsellem94}.} to equation~(61) of \citet{emsellem94} when $b_k=1$ and $M_j=\Upsilon L_k$. As in the semi-isotropic case this formula is very quick to evaluate as it still requires a single numerical quadrature and involves no special functions. All this starting directly from a fit to the galaxy surface-brightness, without the need for numerical deprojection or PSF deconvolution, as required in other approaches. The formula has various possible applications, as it can be used to model realistic galaxies with variable anisotropy or multiple kinematic components, dark-matter, variable stellar $M/L$ and supermassive black holes (representing the point mass with a small gaussian as described in the last paragraph of \refsec{sec:mge_formalism}).

Simple expressions, involving a single quadrature, can be derived for the second velocity moments also for the case of the Keplerian potential (\ref{eq:kepler}) of a black hole, as done in the semi-isotropic case in appendix~A of \citet{emsellem94}. The black hole moments can then be quadratically co-added to the ones in \refeq{eq:second_moment} to obtain the observed velocity moments for the galaxy. We found no advantage in speed or accuracy when performing a separate calculation for the black hole and the galaxy potential. For this reason we will then not give separate expressions for the Keplerian case. Using \refeq{eq:second_moment} with a black hole it is however important to use a quadrature routine which samples the integrand function at a sufficiently high number of initial points, to properly recognize the sharp peak in the integrand near $u=0$. Here we used the vectorized adaptive quadrature algorithm of \citet{shampine08}.

The second moments $\overline{v_{\rm los}^2}$ provided by \refeq{eq:second_moment} are a good approximation for the observed quantity $V^2_{\rm rms}=V^2 + \sigma^2$, where $V$ is the stellar mean velocity and $\sigma$ is the velocity dispersion. In \citet{cappellari06} we used realistic semianalytic dynamical models of galaxies and found that, to extract $\overline{v_{\rm los}^2}$ from the simulated data, one should use a single Gaussian LOSVD and adopt as $V$ and $\sigma$ the mean velocity and dispersion of that Gaussian. Due to the sensitivity of the second moments to the uncertain wings of the LOSVD, this approach is preferable than trying to extract a more complex LOSVD, e.g. by fitting the Gauss-Hermite parametrization \citep{vanDerMarel93,gerhard93} or a fully non-parametric LOSVD, and numerically integrate $\overline{v_{\rm los}^2}$ from that.

\subsubsection{Line-of-sight integration of the velocity first moment}
\label{sec:satoh}

The projected first velocity moments $\overline{v_{\rm los}}\equiv\overline{v_{z'}}$ are given by
\begin{equation}\label{eq:first_moments_projection}
    \Sigma\,\overline{v_{\rm los}} =  \int_{-\infty}^{\infty}
    \nu\overline{v_\phi} \cos \phi \sin i\; \dd z'.
\end{equation}
In this case, the two assumptions we made in \refsec{sec:jeans_solution} are not sufficient any more to provide a unique prediction and therefore additional assumptions are needed. The Jeans equations~(\ref{eq:jeans_beta_R}) and (\ref{eq:jeans_beta_z}) in fact only give a prediction for $\overline{v_{\phi}^2}$ and one has to decide how the second moments separate into the contribution of ordered and random motion, as defined by
\begin{equation}\label{eq:v_phi2}
\overline{v_\phi^2}=\overline{v_\phi}^2 + \sigma_{\phi}^2.
\end{equation}
This need for extra assumptions on the tangential anisotropy is a fundamental limitation of the first-moments equations, and it is the reason why one should fit the more general \refeq{eq:second_moment} to $V_{\rm rms}$, and only subsequently fit any extra parameter of the first moment solution to $V$, instead of simultaneously fitting the Jeans solutions to both $V$ and $\sigma$.

The first moment equations however are very useful to quantify the amount of rotation in galaxies and for this reason have been used in the past (\refsec{sec:review}). One can think of two natural options to parameterize the separation of random and ordered streaming rotation around the symmetry axis. The first option consists of assuming a constant anisotropy, for each Gaussian component, in the $(v_R,v_\phi)$ coordinates, analogously to \refeq{eq:sigma_R}
\begin{equation}\label{eq:no_satoh}
    [\nu\sigma_{\phi}^2]_k = c_k[\nu\overline{v_R^2}]_k,
\end{equation}
which would imply
\begin{equation}
    [\overline{v_\phi}]_k = \left([\overline{v_{\phi}^2}]_k - c_k[\overline{v_R^2}]_k\right)^{1/2}.
\end{equation}
A practical limitation of this approach is that the mean velocity would depend in a non-linear way on the $c_k$ parameter. For this reason nearly all previous authors adopted, in the semi-isotropic case, a second approach, introduced by \cite{satoh80}. This consists of defining a constant $k$ which quantifies how much the model velocity field is scaled with respect to that of the isotropic rotator ($|k|=1$). In this way the first moment can be computed only once, for the isotropic rotator, and is scaled linearly with $k$.

We adopt here the analogue, in our anisotropic case, of \citet{satoh80} approach. It has the added advantage of providing a direct measure of the amount of rotation, which is a more meaningful quantity than the tangential anisotropy. We define for each Gaussian component
\begin{equation}\label{eq:satoh}
    [\overline{v_\phi}]_k = \kappa_k \left([\overline{v_{\phi}^2}]_k - [\overline{v_R^2}]_k\right)^{1/2}.
\end{equation}
Here $\kappa_k=0$ when the $k$-th Gaussian component is not rotating and $|\kappa_k|=1$ when its velocity ellipsoid is a circle in the $(v_R,v_\phi)$ plane. If $b_k=1$ then $\kappa_k$ reduces to \citet{satoh80} parameter and in this case $|\kappa_k|=1$ implies isotropy (the velocity ellipsoid is a sphere everywhere). An upper limit to $|\kappa_k|$ is set by the physical requirement that $\sigma_\phi^2>0$ (equation~[\ref{eq:v_phi2}]).
When summed over all luminous Gaussian components of the MGE model\footnote{As $\overline{v_\phi}^2$ is defined as a quadratic difference of second moments it is {\em not} correct to write $\nu\overline{v_\phi} = \sum_{k=1}^N [\nu\overline{v_\phi}]_k$, as one would normally expect for a first moment.}
\begin{equation}\label{eq:satoh1}
    \nu\overline{v_\phi}^2 = \sum_{k=1}^N [\nu\overline{v_\phi}^2]_k,
\end{equation}
so \refeq{eq:satoh} implies\footnote{To allow for counter-rotating Gaussian components, namely to keep track of the sign of $\kappa_k$ in \refeq{eq:satoh2} and (\ref{eq:first_moments_projection_final}), one should multiply each term of the $k$-summation by ${\rm sgn}(\kappa_k)\equiv\kappa_k/|\kappa_k|$ and then compute ${\rm sgn}(w)\,|w|^{1/2}$, where $w=[\ldots]$ is the expression inside the big square brackets.}
\begin{equation}\label{eq:satoh2}
    \nu\overline{v_\phi} = \left[\nu\sum_{k=1}^N \kappa_k^2 \left([\nu\overline{v_{\phi}^2}]_k - [\nu\overline{v_R^2}]_k\right)\right]^{1/2}.
\end{equation}

Substituting \refeq{eq:satoh2} into \refeq{eq:first_moments_projection}, and using equations~(\ref{eq:sigma_R})--(\ref{eq:vel2_phi}), we obtain the projected first velocity moment of the whole MGE model
\begin{eqnarray}\label{eq:first_moments_projection_final}
    \lefteqn{\Sigma\,\overline{v_{\rm los}}(x',y') =
    2\sqrt{\pi G}\; x'\sin i} \\
    & & \times  \int_{-\infty}^{\infty}
    \left[\nu\int_0^1\!\! \sum_{k=1}^N \sum_{j=1}^M
    \frac{\kappa_k^2 \nu_k q_j \rho_{0j} \mathcal{H}_j(u)\, u^2 \mathcal{D}}
    {1-\mathcal{C} u^2}\dd u \right]^{1/2}\!\!\!\! \dd z'.\nonumber
\end{eqnarray}
When both $b_k=|\kappa_k|=1$ and $M_j=\Upsilon L_k$, this equation reduces to a self-consistent isotropic rotator as in equation~(59) of \citet{emsellem94}. A double quadrature seems unavoidable here, but when $\kappa$ is assumed to be constant for the whole MGE, this integral has to be evaluated only once with $\kappa_k=1$, at the best fitting $(i,\beta_z,\Upsilon)$ parameters previously determined from a fit to the more general second moment \refeq{eq:second_moment}, and then $\overline{v_{\rm los}}$ can be linearly scaled by $\kappa$ to fit the data.\footnote{When the alternative definition for the tangential anisotropy of equation~(\ref{eq:no_satoh}) is used instead of equation~(\ref{eq:satoh}), the term $\mathcal{D}$ in equation~(\ref{eq:first_moments_projection_final}) becomes
$[\mathcal{D} + (1-c_k)\, b_k q_k^2 \sigma_k^2/R^2]$.}

\subsection{Spherical case}

The cylindrically oriented assumption for the shape of the velocity ellipsoid discussed in \refsec{sec:cylindrical_jeans} is likely to be unrealistic for slow-rotator elliptical galaxies. As a class these objects are weakly triaxial, and indeed some of them show clear twists in their kinematical axes \citep{kormendy96,emsellem07,cappellari07,krajnovic08}. The most general and realistic models for these objects are triaxial and orbit- or particle-based \citep{deLorenzi07,vanDenBosch08}. However also in the triaxial approximation a degeneracy in the recovery of the galaxy shape is still generally present so that no unique solution can be obtained. When one is only interested in global galaxy quantities, or to test the results of more general models, it is still  useful to construct simpler and approximate models. The isophotes of the slow rotators are in projection close to circular, especially in their central parts where the kinematics is generally obtained, and this implies they must be intrinsically not far from spherical. This suggests that one can use simple spherical models as a first order approximation to the dynamics of at least some of these objects. The spherical solution of the anisotropic Jeans equations, in the MGE formalism, will be discussed in the following sections.

\subsubsection{General solution}

When the Boltzmann \refeq{eq:boltzmann} is written in spherical coordinates $(r,\theta,\phi)$, by analogy with our derivation of the axisymmetric Jeans equations one can obtain the Jeans equations in spherical symmetry (\citealt{binney82}; equation~[4-54] of BT)
\begin{equation}
\frac{\dd(\nu\overline{v_r^2})}{\dd r}+\frac{2\beta\, \nu\overline{v_r^2}}{r}=-\nu\frac{\dd \Phi}{\dd r},
\end{equation}
where $\overline{v_{\theta}^2}=\overline{v_{\phi}^2}$ for symmetry and we defined $\beta=1-\overline{v_{\theta}^2}/\overline{v_r^2}$.
The solution of this linear first-order differential equation with constant anisotropy $\beta$ and the boundary condition $\nu\overline{v_r^2}=0$ as $r\rightarrow\infty$ is given by \citep[e.g.][]{vanDerMarel94}
\begin{eqnarray}\label{eq:spherical_jeans}
    \lefteqn{\nu\overline{v_r^2}(r) = r^{-2\beta }
    \int_r^{\infty}\nu(u)
    \frac{\dd\Phi(u)}{\dd u}u^{2\beta}\dd u\nonumber}\\
    && = G\, r^{-2\beta}
    \int _r^{\infty }\frac{\nu(u)M(u)}{u^{2-2\beta}}\dd u,
\end{eqnarray}
considering that $\dd\Phi/\dd r=G M/r^2$. After projection along the line-of-sight $z'$, we obtain the observed second velocity moment (see equation~[4-60] of BT)
\begin{eqnarray}
\lefteqn{\Sigma \overline{v_{\rm los}^2}(R)=2\int _R^{\infty}
\left(1-\beta \frac{R^2}{r^2}\right)
\frac{\nu\overline{v_r^2}(r)r }{\sqrt{r^2-R^2}}\,\dd r}\\
&&=2G\int_R^{\infty }
\left[\frac{r^{-1-2 \beta } \left(r^2-R^2 \beta \right)}{\sqrt{r^2-R^2}}
\int _r^{\infty }\frac{\nu(u)M(u)}{u^{2-2\beta}}\dd u\right] \, \dd r\nonumber,
\end{eqnarray}
where $R$ is the projected radius, measured from the galaxy center. Integrating by parts the double integral can be reduced to a single quadrature, involving special functions
\begin{equation}\label{eq:sph_jeans}
\Sigma \overline{v_{\rm los}^2}(R) = 2G\int_R^{\infty}
\frac{\mathcal{F}(r)\nu(r)M(r)}{r^{2-2\beta}}\dd r,
\end{equation}
where
\begin{eqnarray}\label{eq:spherical_anisotropic_final}
\lefteqn{\mathcal{F}(r)=\frac{R^{1-2\beta }}{2}\left[\beta\, B_w\left(\beta+\frac{1}{2} ,\frac{1}{2}\right)-B_w\left(\beta-\frac{1}{2} ,\frac{1}{2}\right)\right.}\nonumber\\
&&\left.+\frac{\sqrt{\pi }\,(\frac{3}{2}-\beta )\,\Gamma\left(\beta-\frac{1}{2} \right)}{\Gamma(\beta )}\right],
\end{eqnarray}
with $w=(R/r)^2$, $\Gamma$ is the Gamma function and $B_w$ is the incomplete Beta function (equation~[6.6.1] of \citealt{abramowitz65}). In the isotropic limit
\begin{equation}
\lim_{\beta\rightarrow0}\mathcal{F}(r)=\sqrt{r^2-R^2},
\end{equation}
and \refeq{eq:sph_jeans} reduces to the spherical isotropic form of equation~(29) of \citet{tremaine94}. Some numerical implementations provide the incomplete Beta function only for positive arguments \citep[e.g.][]{press92}, for negative values ($\beta<1/2$) one can use its analytic continuation via Gauss's Hypergeometric function $_2F_1$ (equation~[6.6.8] of \citealt{abramowitz65})
\begin{equation}
B_w(a,b)=\frac{w^a}{a}\, _2F_1(a,1-b;a+1;w),
\end{equation}
for which efficient routines\footnote{Available from http://jin.ece.uiuc.edu/routines/routines.html} exist \citep{shanjie96}. For $\beta=\pm1/2$, the $B_w$ function is divergent, but for all practical purposes we found it is sufficient to perturb the $\beta$ value by a negligible amount to avoid the singularity.

\subsubsection{MGE spherical Jeans solution}
\label{sec:mge_spherical}

We proceed as in \refsec{sec:mge_solution} to derive an explicit solution for the spherical Jeans equation using the MGE parametrization for both the luminosity density and the total density. The expressions for the surface brightness, luminosity density and total density are as in \refeq{eq:surf}, (\ref{eq:dens}) and (\ref{eq:mass}), with $q_j=q_k=1$ \citep{bendinelli91}. For a single Gaussian component they read
\begin{equation}\label{eq:surf_sph}
\Sigma_k(R) =
{\frac{L_k}{2\pi\sigma^2_k}
\exp\left(-\frac{R^2}{2\sigma^2_k}\right)
},
\end{equation}
\begin{equation}\label{eq:dens_sph}
\nu_k(r) =
\frac{L_k}{(\sqrt{2\pi}\, \sigma_k)^3}
\exp\left(-\frac{r^2}{2\sigma_k^2}\right)
\end{equation}
\begin{equation}\label{eq:mass_sph}
\rho_j(r) =
\frac{M_j}{(\sqrt{2\pi}\, \sigma_j)^3}
\exp\left(-\frac{r^2}{2\sigma_j^2}\right).
\end{equation}
The mass of a Gaussian contained within the spherical radius $r$ is
\begin{equation}\label{eq:massr}
M_j(r)=M_j \left[{\rm erf}\left(\frac{r}{\sqrt{2}\, \sigma_j}\right)
-\frac{\sqrt{\frac{2}{\pi}}\; r}{\sigma_j}
\exp\left(-\frac{r^2}{2 \sigma_j^2}\right)
\right],
\end{equation}
with ${\rm erf}(x)$ the error function (equation~[7.1.1] of \citealt{abramowitz65}). Using \refeq{eq:sph_jeans} and summing over all Gaussian components with \refeq{eq:sum_vz2}, we obtain the projected second velocity moment for the whole MGE model
\begin{equation}\label{eq:sph_jeans_mge}
\Sigma \overline{v_{\rm los}^2}(R) = 2G\!\!\int_R^{\infty}
\sum_{k=1}^N \frac{\mathcal{F}_k(r)\,\nu_k(r)}{r^{2-2\beta_k}}\!
\left[M_\bullet+\sum_{j=1}^M M_j(r)\right]
\dd r,
\end{equation}
where $\nu_k(r)$ and $M_j(r)$ are given by \refeq{eq:dens_sph} and (\ref{eq:massr}) respectively, and $\mathcal{F}_k(r)$ is obtained by replacing the $\beta$ parameter in \refeq{eq:spherical_anisotropic_final} with the anisotropy $\beta_k$ of each luminous Gaussian component. We explicitly included the mass $M_\bullet$ of a central supermassive black hole. As in the axisymmetric case, this formula involves a single numerical quadrature. It can be used to model nearly spherical objects with a variable anisotropy profile, variable stellar $M/L$, a supermassive black hole and dark matter.

An approximate way to construct a model with a certain smooth radial profile of anisotropy $\beta(r)$, consists of defining $\beta_k=\beta(\sigma_k)$ for each MGE Gaussian with dispersion $\sigma_k$. This simple method works because in a spherical MGE each $k$-th Gaussian produces a significant contribution to the total luminosity density only near its $\sigma_k$ (\citealt{cappellari02mge}; see also equation~[\ref{eq:var_beta}]).

\subsection{Availability}

The Jeans Anisotropic MGE (\textsc{JAM}) package of \textsc{IDL}\footnote{http://www.ittvis.com/idl/} procedures, providing a reference implementation for the equations described in this section, together with other routines to evaluate auxiliary quantities, like the circular velocity, from MGE models, is available online from http://www-astro.physics.ox.ac.uk/$\sim$mxc/idl/.

\section{Applications and tests}
\label{sec:tests}

The formalism derived in Sections~\ref{sec:los_mu2} and \ref{sec:satoh} generalizes to anisotropy the widely used semi-isotropic formalism for the solution of the Jeans equations while maintaining its simplicity. It was motivated by our modeling results on the orbital distribution of a sample of real galaxies \citep{cappellari07}. However the method would be of limited usefulness if it did not describe the dynamics of fast-rotator early-type galaxies well. In this section we show that the simple assumptions we made provide a remarkable good description of the main features of the kinematics\footnote{Available from http://www.strw.leidenuniv.nl/sauron/.} \citep{emsellem04} of the fast rotator early-type galaxies, as observed with the \sauron\ integral-field spectrograph \citep{bacon01} as part of the \sauron\ survey \citep{dezeeuw02}.

\subsection{Comparison with more general models}

\subsubsection{Anisotropy comparison}
\label{sec:comp_schw}

\begin{figure*}
\centering
  \includegraphics[height=0.39\textheight]{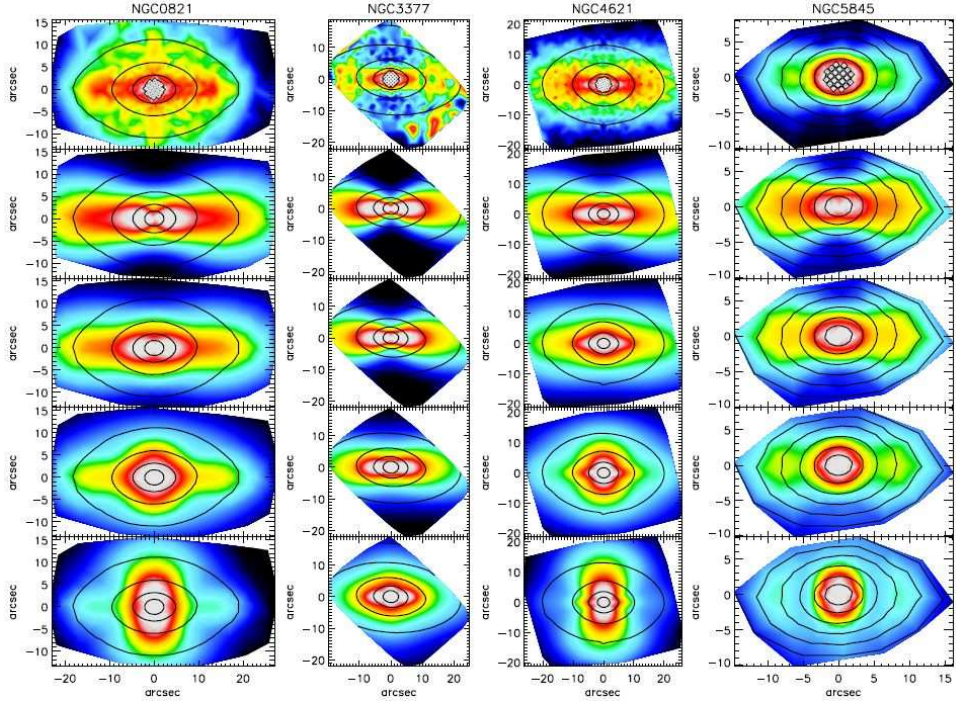}
  \caption{Data-model comparison for the second velocity moments of edge-on galaxies. From left to right, the columns show the galaxies NGC~821, NGC~3377, NGC~4621 and NGC~5845. The top row visualizes the bi-symmetrized \sauron\ observations for $V_{\rm rms}=\sqrt{V^2+\sigma^2}$. The subsequent rows show the model predictions for $(\overline{v_{\rm los}^2})^{1/2}$ as given by \refeq{eq:second_moment} for $\beta_z=0$ (semi-isotropic model; second row) and anisotropic models with $\beta_z=0.1$, 0.2 and 0.3 respectively. The color levels are the same for the five panels in each column. NGC~4621 appears well reproduced by a weakly anisotropic model $\beta_z\approx0.1$. NGC~812 and NGC~5845 are well described by a model with $\beta_z\approx0.2$, while only the strongest anisotropy $\beta_z\approx0.3$ can qualitatively describe the observed shallow gradient of $V_{\rm rms}$ along the projected minor axis of NGC~3377. These estimates are more accurately quantified in \reffig{fig:chi2_beta}.
  }\label{fig:models_beta}
\end{figure*}

In this section we compare the anisotropy derived from the Jeans models, with the results for the global anisotropy obtained with the more general \citet{schwarzschild79} method on the same galaxies and from the same data in \citet{cappellari07}. We use \refeq{eq:second_moment}, with constant anisotropy $b_k=b$ and constant mass-to-light ratio $\Upsilon$ for the whole MGE model, to predict the velocity second moments. Following standard practice we parameterize the anisotropy with the variable $\beta_z\equiv1-\overline{v_{z}^2}/\overline{v_{R}^2}=1-1/b$. The problem becomes a function of two nonlinear variables $(i,\beta_z)$ and of the scaling factor $\Upsilon$. To further limit the number of free variables in the model, and to eliminate the degeneracy in the deprojection of the surface brightness, we select for the comparison the galaxies for which the photometry already constrains the inclination to be close to edge-on ($i=90^\circ$). For this we select from the MGE models in Table~B1-B2 of \citet{cappellari06}, the fast-rotator galaxies for which the flattest Gaussian has $q'<0.35$, which forces a strict limit on the inclination $i>70^\circ$. We further exclude the galaxy NGC~4550, due to the presence of two counter-rotating stellar disks, which complicate the interpretation of the models, and NGC~4526, which has a strong dust disk affecting the observed stellar kinematics. This selection leads to the four galaxies NGC~821, NGC~3377, NGC~4621 and NGC~5845.

We evaluated the MGE model predictions given by \refeq{eq:second_moment}, adopting $i=90$ as in \citet{cappellari07}, for a direct comparison with the anisotropy determinations. The model predictions were convolved with the PSF and integrated over the pixels, before comparison with the observables, as described in Appendix~\ref{sec:psf}.  For each galaxy we computed the model predictions for $(\overline{v_{\rm los}^2})^{1/2}$ inside each Voronoi bin on the sky, at different anisotropy values $\beta_z$. At every anisotropy the best fitting mass-to-light ratio $\Upsilon$ is obtained from the simple scaling relation $\Upsilon\propto\overline{v_{\rm los}^2}$ as a linear least-squares fit. The best-fitting scaling factor for the model velocities is $s=\cos{\alpha}\,|\mathbf{d}|/|\mathbf{m}|$, where $\alpha$ is the cosine of the angle between the data $\mathbf{d}$ and model $\mathbf{m}$ vectors, which implies:\footnote{We corrected a typo in the corresponding equation~(2) of \citet{cappellari06}.}
\begin{equation}\label{eq:ml_fit}
\Upsilon=\left(\frac{\mathbf{d}\cdot\mathbf{m}}
{\mathbf{m}\cdot\mathbf{m}}\right)^2,
\end{equation}
where the vectors $\mathbf{d}$ and $\mathbf{m}$ have elements
$d_n=[V_{\rm rms}/\Delta V_{\rm rms}]_n$ and $m_n=[(\overline{v_{\rm los}^2})^{1/2}/\Delta V_{\rm rms}]_n$, with $[V_{\rm rms}]_n\equiv\sqrt{V_n^2+\sigma_n^2}$ the measured values for the $P$ Voronoi bins from \citet{emsellem04} and $[\Delta V_{\rm rms}]_n$ the corresponding errors, while $[(\overline{v_{\rm los}^2})^{1/2}]_n$ are the model predictions for $\Upsilon=1$. A qualitative comparison between the observed velocity second moments and the model ones, for different values of the anisotropy parameter $\beta_z$ is shown in \reffig{fig:models_beta}.

\begin{figure}
\centering
  \plotone{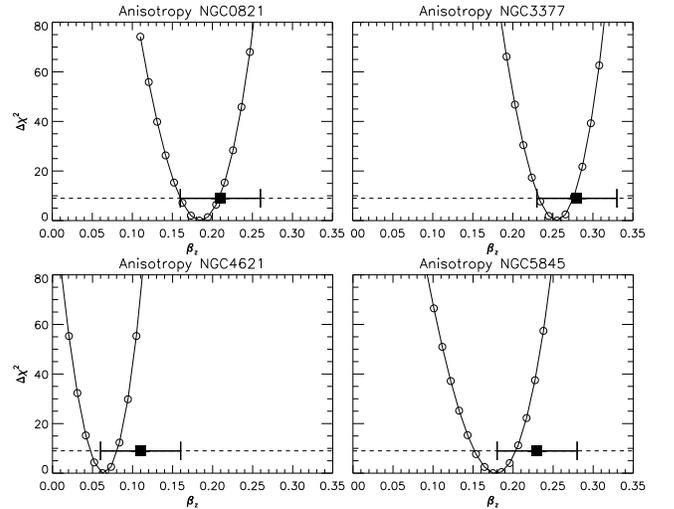}
  \caption{Best-fitting Jeans anisotropy. $\Delta\chi^2=\chi^2-\chi^2_{\rm min}$, describing the agreement between the data and the models, is plotted against the anisotropy parameter $\beta_z$. The four panels show the galaxies NGC~821, NGC~3377, NGC~4621 and NGC~5845. The solid line with open circles represent the $\Delta\chi^2$ of the Jeans models. The dashed horizontal line indicates the level $\Delta\chi^2=9$, which corresponds to the $3\sigma$ confidence level for one degree-of-freedom. The filled square with error bars shows the value of the global anisotropy measured using Schwarzschild models in \citet{cappellari07}.}\label{fig:chi2_beta}
\end{figure}

To ease the visual comparison, given that the models are bi-symmetric by construction, the observations have been bi-symmetrized with respect to the kinematical major axis PA$_{\rm kin}$ given in \citet{cappellari07}. The actual fits and $\chi^2$ calculations are always performed using the original data and errors. It is clear that the isotropic models ($\beta_z=0$) do not provide a good description for the observed second velocity moments of any of the four galaxies\footnote{The difference between the isotropic models of NGC~821 and NGC~3377, and the observations, has already been shown in figure~A1 of \citet{cappellari07}.}, the disagreement being strongest for NGC~3377 and weakest for NGC~4621. Using the models presented in this paper we can now move away from the isotropic assumption and quantify that the model of NGC~4621 requires a small amount of anisotropy to appear qualitatively like the data: any $\beta_z\ga0.1$ produces in fact two vertical `lobes' along the minor axis, which are not observed. Both NGC~821 and NGC~5845 require a more significant anisotropy $\beta_z\approx0.2$ to reduce the amount of rotation (high $[\overline{v_{\rm los}^2}]^{1/2}$) along the major axis to the level of the observations. A larger $\beta_z\approx0.3$ is strongly excluded by the data, as the model immediately produces a significant elongation along the minor axis, which is not observed. Finally, in the case of NGC~3377, only the most extreme anisotropy shown ($\beta_z\approx0.3$) is able to reduce the gradient of $(\overline{v_{\rm los}^2})^{1/2}$ along the minor axis to the shallow level of the observations.

To quantify more precisely the findings of the previous paragraph, in \reffig{fig:chi2_beta} we show the plots of $\Delta\chi^2=\chi^2-\chi^2_{\rm min}$, describing the agreement between the data and the models, as a function of $\beta_z$. We also show, as a filled square with error bars, the value of the global anisotropy we measured using more general axisymmetric Schwarzschild's models in \citet{cappellari07}. The $\Delta\chi^2$ plots confirm the visual impression from the maps of \reffig{fig:models_beta} and the fact that the Jeans models give a tight constrain to the anisotropy. The small size of the Jeans formal errors is of course illusory, as it is due to the restrictive model assumptions. However the comparison with the more general results show that the simple anisotropic Jeans models are able to recover the anisotropy within the errors. There seems to be a systematic difference of $\Delta\beta_z\approx0.05$ between the anisotropic Jeans and the Schwarzschild determinations, with the Jeans results being lower. This difference is at the level of the errors in the Schwarzschild models (see section 4.3 in \citealt{cappellari07}), and not particularly surprising given the radically different modeling approaches adopted. If real, it may be due to the fact that the Jeans models force a constant anisotropy everywhere in the galaxy model, while the Schwarzschild ones allow for a more realistic and more spherical velocity ellipsoid at intermediate latitudes ($\theta\approx45$; see \reffig{fig:coordinates}). The Jeans models may need a lower anisotropy to compensate the mismatch with the data at these intermediate latitudes.

\subsubsection{$M/L$ comparison}
\label{sec:ml_comparison}

\begin{figure}
  \plotone{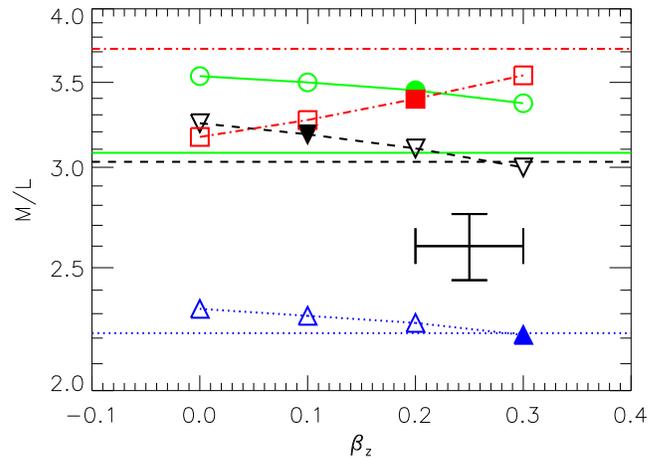}
  \caption{$M/L$ versus anisotropy. The different symbols, connected by lines, show the variation with anisotropy $\beta_z$ in the dynamical $M/L$ ($I$-band) for each of the Jeans models shown in \reffig{fig:models_beta}. Also shown for reference, with horizontal lines using the same style and color, is the $(M/L)_{\rm Schw}$, determined from the same data in \citet{cappellari06}, using three-integral Schwarzschild models. The symbols represent NGC~821 (green circles), NGC~3377 (blue upward triangles), NGC~4621 (black downward triangles) and NGC~5845 (red squares). The filled symbols correspond to the best-fitting $\beta_z$ from \reffig{fig:chi2_beta}. The large cross shows the characteristic error in $\beta_z$ and $(M/L)_{\rm Schw}$.}\label{fig:ml_variation}
\end{figure}

When studying the stellar population or the dark-matter content of galaxies, one is interested more in the $M/L$ determination than in the anisotropy. It is important to test whether an error in the anisotropy can lead to a significant error in $M/L$.  For this in \reffig{fig:ml_variation} we show the variation in the dynamical $M/L$ ($I$-band) for each of the models shown in \reffig{fig:models_beta}. Also shown for reference is the $(M/L)_{\rm Schw}$ determined from the same \sauron\ data in \citet{cappellari06}, using three integral Schwarzschild models. The comparison shows that the $M/L$ varies very little for the ranges of anisotropy observed in fast rotators early-type galaxies. In all cases the Jeans $M/L$ at the best-fitting $\beta_z$ represents an improvement over the $M/L$ determined from semi-isotropic models (already given in Table~1 of \citealt{cappellari06}). With the exception of NGC~5845, the improvement is generally at the level of the measurement errors. The latter galaxy is an interesting case on its own: it has an small size with $\re\approx4\farcs6$ \citep{cappellari06} and constitute a unique example of an elliptical galaxy with integral-field stellar kinematics out to $\approx3\re$. Our standard self-consistent model still well reproduces the \sauron\ kinematics without the need to invoke a dark matter halo.

The small dependence of $M/L$ on the anisotropy explains the fact that, in a comparison between the semi-isotropic $(M/L)_{\rm Jeans}$ and the three-integral $(M/L)_{\rm Schw}$ for a sample of 25 early-type galaxies (including the ones in \reffig{fig:ml_variation}), we found an excellent agreement between the two determinations. We did not detect any significant bias in the Jeans values and the differences could be explained by random errors at the 6 per cent rms level \citep[their fig.~7]{cappellari06}.

\subsection{First and second velocity moments model examples}
\label{sec:velocity_examples}

\begin{figure*}
\centering
  \includegraphics[height=0.3\textheight]{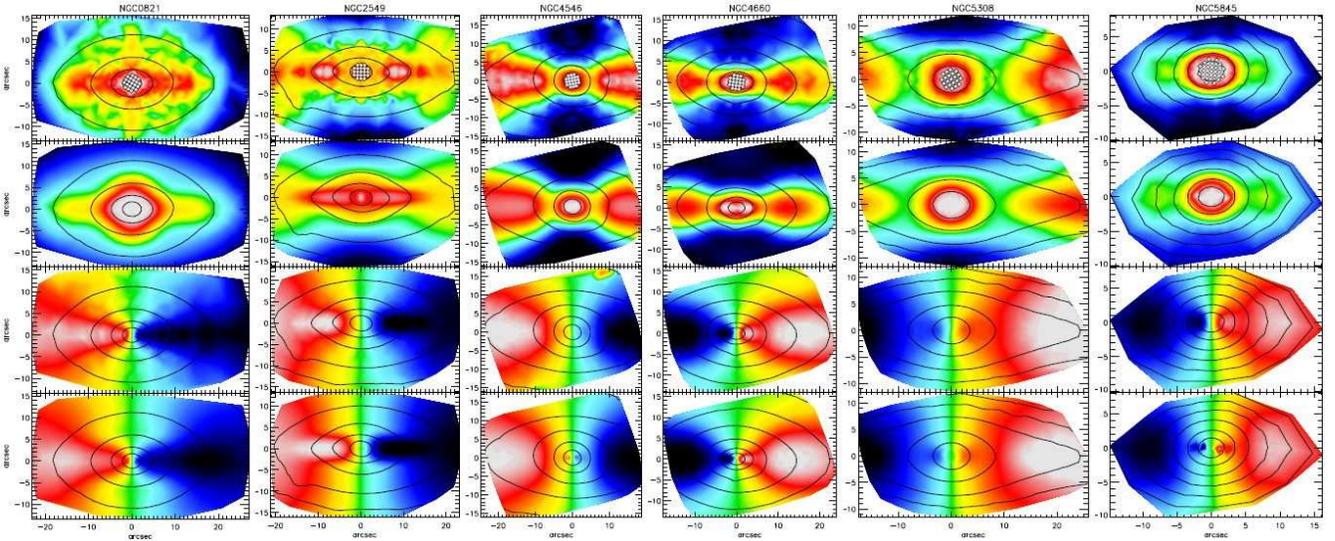}
  \caption{Data-model comparison for three E and three S0 fast-rotator galaxies. From left to right, the columns show the fast-rotators NGC~821 (E; $i=82^\circ, \beta_z=0.20, \kappa=0.75$), NGC~2549 (S0; $i=90^\circ, \beta_z=0.17, \kappa=0.99$), NGC~4546 (S0; $i=71^\circ, \beta_z=0.05, \kappa=1.00$), NGC~4660 (E; $i=67^\circ, \beta_z=0.23, \kappa=0.93$), NGC~5308 (S0; $i=87^\circ, \beta_z=0.28, \kappa=1.02$) and NGC~5845 (E; $i=75^\circ, \beta_z=0.18, \kappa=0.95$). The top row visualizes the bi-symmetrized \sauron\ observations for $V_{\rm rms}=\sqrt{V^2+\sigma^2}$. The second row show the best-fitting model predictions for $(\overline{v_{\rm los}^2})^{1/2}$ as given by \refeq{eq:second_moment}. The third row presents the observed \sauron\ mean velocity $V$. The bottom row shows the best fitting model first velocity moment $\overline{v_{\rm los}}$ as given by \refeq{eq:first_moments_projection_final}. The color levels are the same for data and model. These galaxies are all constrained by the photometry to be quite close to edge-on, so the models can vary essentially only the {\em single} parameter $\beta_z$ to fit the shape of the observed first and second moments. The kinematics varies widely for different galaxies, yet this single parameter is sufficient to correctly predict the main features of a pair of two-dimensional functions ($V_{\rm rms}$ and $V$), once the observed surface brightness is given.
  }\label{fig:models_velocity}
\end{figure*}

In the previous section we showed that the anisotropic Jeans formalism introduced in this paper provides a good qualitative description of the integral-field second velocity moments observed with \sauron. We also showed that for edge-on galaxies the recovered anisotropy agrees well with the one derived from more general models. Here we present additional examples of model fits to the second moments and we show that, using the simple \citet{satoh80} assumption for the splitting of the random and ordered rotation (\refsec{sec:satoh}), a remarkably good prediction of the galaxies mean velocity can be obtained as well, in many cases.

In \reffig{fig:models_velocity} the data-model comparison is presented for both the first and second velocity moments. The model fits were determined with the following procedure:
\begin{enumerate}
\item For each pair of the nonlinear model parameters $(i,\beta_z)$ we evaluated the predicted second moments $(\overline{v_{\rm los}^2})^{1/2}$ using \refeq{eq:second_moment}, with $b_k=1/(1-\beta_z)$ and $M_j=L_k$.;
\item The model predictions were scaled by the best-fitting mass-to-light ratio $\Upsilon$ via \refeq{eq:ml_fit} and the $\chi^2$, describing he agreement of data and models was determined;
\item Step (i)--(ii) were repeated for a grid of $(i,\beta_z)$ parameters and the best fitting value was found. In the grid, the inclination was sampled at equally spaced intervals in the intrinsic axial ratio of the flattest MGE component, while $\beta_z$ was sampled linearly;
\item The Gaussians describing the mass of the MGE (equation~[\ref{eq:mass}]) were scaled by the best fitting $\Upsilon$ and a prediction for the first velocity moment $\overline{v_{\rm los}}$ was computed with \refeq{eq:first_moments_projection_final}, with $\kappa_k=1$;
\item The model velocity field $\overline{v_{\rm los}}$ was scaled by the factor
    \begin{equation}
    \kappa=
    \frac{\sum_{n=1}^P F_n |x'_n V_n|}
         {\sum_{n=1}^P F_n |x'_n [\overline{v_{\rm los}}]_n|},
    \end{equation}
    where $P$ is the number of Voronoi bins in the observed \sauron\ data and $F_n$ is the corresponding flux. $V_n$ and $[\overline{v_{\rm los}}]_n$ are the observed and model velocities respectively, and $x'_n$ are the bin coordinates (the $x'$ axis being along the projected major axis).
\end{enumerate}
The last step ensures that the model has the same projected angular momentum as the observed galaxy, within the observed region. Doing a normal least-squares fit would heavily underestimate the amount of rotation of the model with respect to the data, in cases where the galaxy contains counter-rotating stellar components, which are clearly excluded by the simple assumption of a constant $\kappa$ factor for the whole MGE model. The angular momentum of the model could be computed by numerically integrating $L_z=\nu R\overline{v_{\phi}}$ over the galaxy volume using \refeq{eq:vel2_phi}. More useful may be to use the fitted galaxy inclination to deprojected the {\em observed} galaxy stellar angular momentum per unit mass \citep{emsellem07}.
To model kinematically distinct stellar components one could allow for different $\kappa_k$ components in \refeq{eq:first_moments_projection_final}, and perform a standard least-squares fit as in \refeq{eq:ml_fit}.

\begin{figure*}
\centering
  \includegraphics[width=0.8\textwidth]{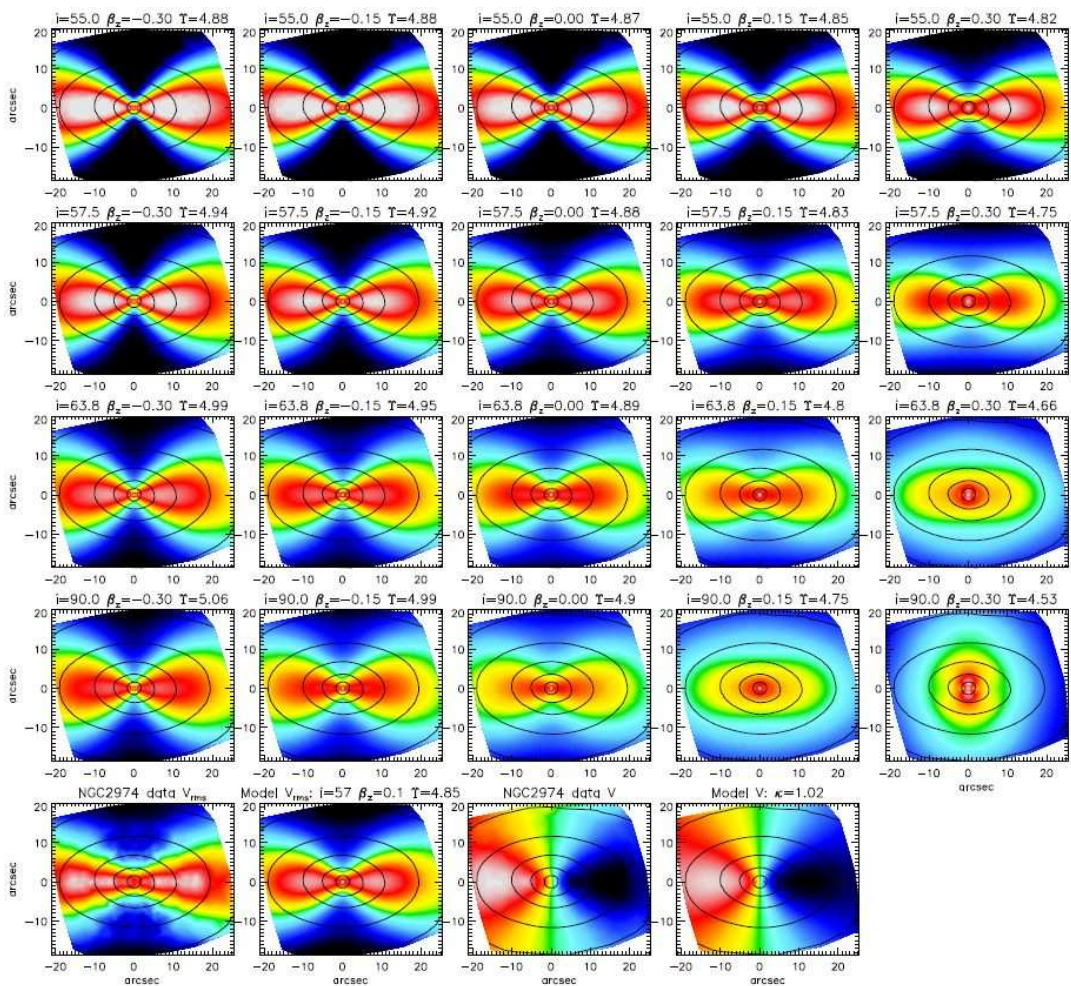}
  \caption{Inclination and anisotropy variation for NGC~2974. Each of the five columns show the model predictions for the second velocity moment $(\overline{v_{\rm los}^2})^{1/2}$ computed at different values of $\beta_z=-0.3,-0.15,\ldots,0.3$. Different rows correspond to different inclinations $i$, equally spaced in the axial ratio of the flattest Gaussian in the MGE model of NGC~2974. The model mass-to-light ratio $\Upsilon$ was optimized for each fit using \refeq{eq:ml_fit}. The values for the input parameters $(\beta_z,i)$ and for the best fitting $\Upsilon$ are printed on top of each panel. The bottom row shows (i) the bi-symmetrized \sauron\ data $V_{\rm rms}=\sqrt{V^2+\sigma^2}$; (ii) the best fitting model for $(\overline{v_{\rm los}^2})^{1/2}$; (iii) the bi-symmetrized \sauron\ data for $V$ and (iv) the best-fitting model for $\overline{v_{\rm los}}$. The colormap has the same scaling for all the panels.}
  \label{fig:models_beta_inc}
\end{figure*}

For the examples of \reffig{fig:models_velocity} we selected three ellipticals and three lenticulars fast-rotator showing a range of kinematical properties. For all the galaxies the MGE models\footnote{The MGE parameters for NGC~2549, NGC~4546 and NGC~5308 are taken from Scott et al. (in preparation), while the MGE models for the other galaxies are given in \citet{cappellari06}.} were fitted to the observed photometry using the public software$^4$ of \citet{cappellari02mge}. Although some differences between the data and models are visible, the $\chi^2$ per degree-of-freedom in the fit presented, computed from original non-symmetrized data, is close to one for all the fits. This indicates that the models are generally consistent with the data, within the measurements errors, which is remarkable for what is essentially a one parameter model! A more general statistical investigation of the kinematic parameters for a larger sample of galaxies goes beyond the scope of the present paper and will be presented elsewhere. Here we note that for the galaxies presented in \reffig{fig:models_velocity} and \ref{fig:models_beta_inc} the parameter $\kappa\approx1$ within a few percent accuracy (except for NGC~821). This appears a general characteristics also of the other fast-rotators we modeled. It implies $\sigma_{\phi}^2\approx\overline{v_{R}^2}$ and confirms the results of more general models. It indicates that the velocity ellipsoid of fast-rotator early-type galaxies tends to be oblate \citep[their fig.~2]{cappellari07}.

Interestingly \reffig{fig:models_velocity} shows that even the complex double-hump structure observed in the velocity fields of NGC~2549, NGC~4660 or NGC~5845, which seems to imply a complex dynamical structure in the galaxy, can be well reproduced by these one-parameter models, once an accurate MGE parametrization for the surface brightness is given. The structures are explained by the presence of thin stellar disks, already visible in the photometry as strongly disky isophotes. There is no evidence for a significantly different anisotropy between the bulge and the disk of these galaxies, as a constant anisotropy well reproduces the observations (within 1\re). These examples show that a great deal of information on the kinematical structure of the fast-rotators is contained in the photometry alone!

\subsection{Recovery of the galaxy inclination}

We have shown in the previous sections that the anisotropic Jeans axisymmetric models presented in this paper, varying essentially only one free parameter $\beta_z$, provide a good qualitative description of the observed shape of the first and second velocity moments of fast-rotator early-type galaxies, when they are already constrained by the photometry to be close to edge-on. We also confirmed that the fast-rotators are characterized by a velocity ellipsoid which is flattened in the $z$ direction so that the anisotropy $\beta_z\ga0$  \citep{cappellari07}. In this section we study the variations of the models when the inclination is allowed to vary.

\subsubsection{The inclination degeneracy}

In \citet{krajnovic05} we showed, using real data and analytic tests, that there appears to be a degeneracy in the recovery of the galaxy inclination (or its corresponding shape) using general three-integral axisymmetric models. This is true even in the case of state-of-the-art integral-field data, and assuming the galaxy potential is accurately known at every inclination and can be uniquely recovered from the surface brightness. This result was confirmed for a larger sample of galaxies in \citet{cappellari06}, and in the oblate limit, but using a triaxial modeling code, by \citet{vanDenBosch09}. In a realistic situation both the galaxy potential and the stellar luminosity density are only approximately known, due to the important intrinsic degeneracy in the deprojection of the surface brightness, due to the possible presence of bars, and due to the likely contribution of a (small) fraction of dark-matter. Moreover the kinematical data are always affected by low-level systematics, which are difficult to control and quantify. The dynamical models are also affected by numerical approximations and discretisation effects. All this makes it unlikely that any inclination derived from dynamical models of the stellar kinematics can be trusted, unless further assumptions are made.
In this section we explore whether the anisotropic Jeans models, combined with realistic assumptions on the galaxy anisotropy, can be used to recover the galaxy inclination.

\subsubsection{Inclination recovery with the Jeans models}
\label{sec:inc_recovery}

For our experiments on the recovery of the inclination we selected the same four galaxies with \sauron\ integral-field kinematics that we presented for the same reason in appendix~A of \citet{cappellari06}. These galaxies have a photometry which, under the axisymmetric assumption, can be deprojected for a wide range of inclinations. Most importantly they possess disks of gas or dust, which allow the inclination to be determined independently from the stellar dynamics. In that paper we found that the inclination derived using semi-isotropic models appeared in agreement with the one from the dust disks. Here we try to understand the reason for the good agreement.

Our prototypical test case is the fast-rotator E4 galaxy NGC~2974. The inclination of this galaxy was studied extensively by a number of authors using the kinematics of the gas \citep[e.g.][]{weijmans08}. From accurate models of the \sauron\ gas kinematics in the regions where we have our stellar kinematics \citet{krajnovic05} find a best fitting inclination $i=60\pm3$. A sequence of model predictions for the velocity second moments $(\overline{v_{\rm los}^2})^{1/2}$ as a function of anisotropy $\beta_z$ and inclination $i$ is shown in \reffig{fig:models_beta_inc}. A strong similarity is evident between the effect on the models of the anisotropy and of the inclination. Another general feature one can see from the figure is that the second moments are more weakly sensitive to an anisotropy variation when the galaxy is intrinsically flat ($i=55^\circ, q\approx0.32$) than when it is rounder ($i=90^\circ, q\approx0.63$). A thin disk can be more strongly anisotropic without producing the vertical elongation of the second moments (see $i=90^\circ, \beta_z=0.3$), which is never observed in real galaxies.

\begin{figure}
\centering
  \plotone{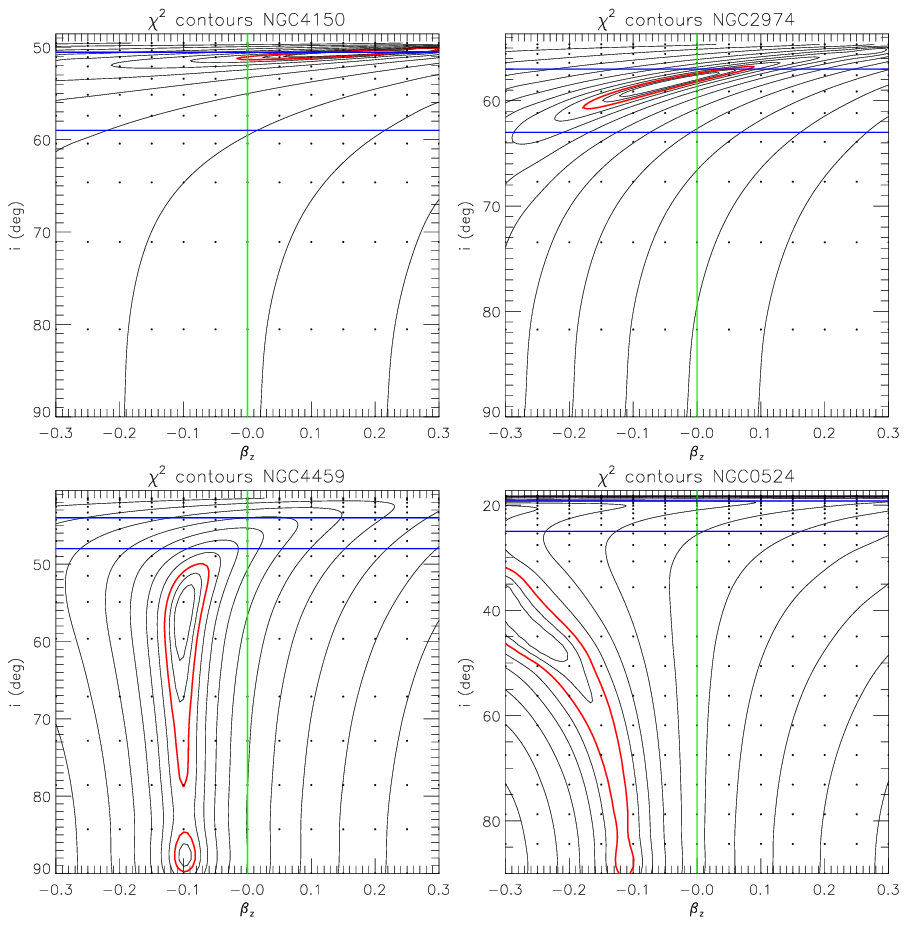}
  \caption{The inclination-anisotropy degeneracy. Contours of $\Delta\chi^2=\chi^2-\chi^2_{\rm min}$, describing the agreement between the data and the anisotropic Jeans models, are plotted as a function of the anisotropy parameter $\beta_z$ and the galaxy inclination $i$ in degrees.
  The lowest levels correspond to the formal 1, 2 and 3$\sigma$ (thick red line with $\Delta\chi^2=11.8$) confidence levels for two degrees of freedom. Additional contours are characterized by a factor of 2 increase in $\Delta\chi^2$.  The two horizontal blue solid lines enclose the region of acceptable inclinations as determined from the gas kinematics or dust geometry (appendix~A of \citealt{cappellari06}). The vertical solid green line indicates isotropy ($\beta_z=0$).
  The different panels correspond to NGC~4150, NGC~2974, NGC~4459 and NGC~524 respectively. There is a clear degeneracy between inclination and $\beta_z$ anisotropy. However if one enforces the observationally-motivated constraint $\beta_z>0.05$, the recovered inclination always lies within the error bars of the gas/dust determinations.}
  \label{fig:chi2_beta_inc}
\end{figure}

The observed behavior can be qualitatively understood as follows: A more face-on view makes the model intrinsically flatter and disk-like, causing the second moments to approach the $|\cos\phi|$ behavior along the galaxy isophotes as one would expect for a thin rotating disk, with a peak on the projected major axis and a strong minimum on the minor axis. A similar increase of the second moments on the projected major axis, with respect to the minor axis, can be produced by reducing the fraction of stars on radial orbits (lowering $\beta_z$), thus correspondingly increasing the stars on tangential orbits. Due to projection these in fact produce zero motion along the minor axis and show a peak on the major axis. For example in \reffig{fig:models_beta_inc} the model predictions with $(i,\beta_z)=(63.8^\circ,-0.3)$ are very similar to the ones for $(i,\beta_z)=(57.5^\circ,0.0)$. The contours of the $\chi^2$, describing the agreement between the Jeans models and the observed \sauron\ $V_{\rm rms}$ are shown in \reffig{fig:chi2_beta_inc}. Within a 3$\sigma$ confidence level any inclination between $i=56^\circ$ and $61^\circ$, and correspondingly any anisotropy between $\beta_z=-0.2$ and 0.1 is equally consistent with the data.

In \citet[their fig.~2]{cappellari07} we found that all the fast-rotators have anisotropy $\beta_z\ga0.05$. For the nearly edge-on galaxies we confirmed this determination using Jeans models in \refsec{sec:comp_schw} and \refsec{sec:velocity_examples}. Forcing this observationally-motivated constraint, the inclination of NGC~2974 becomes tightly constrained to the range $i=56^\circ$ to $58^\circ$. The qualitative behavior seen in NGC~2974 is representative of what we observed for the 48 galaxies we modeled using this method from the \sauron\ sample (Scott et al.\ in preparation). In \reffig{fig:chi2_beta_inc} we show the $\chi^2$ contours for the three additional galaxies. In the case of NGC~4150 the inclination is well constrained by the data independently of anisotropy, while in the case of NGC~4459 and NGC~524 the best fitting location has $\beta_z<0$ and the inclination is heavily dependent on the assumed anisotropy. For these two galaxies the un-physical location of the $\chi^2$ minimum is likely due to the degeneracy in the deprojection of the surface brightness (\refsec{sec:mge_formalism}). These two galaxies have a quite low inclination and the deprojected mass distribution is probably significantly in error even at the true inclination, which is then formally excluded by the data. Like all the fast-rotators, these galaxies likely possess stellar disks, which cannot be detected at these low inclinations \citep{rix90,magorrian99}. This would explain the general fact that we tend to better reproduce the kinematics of the fast-rotators which are close to edge-on than the kinematics of the more face-on ones.
Nonetheless, in all the four cases, if one enforces the constrain $\beta_z\ga0.05$, the recovered inclination lies within the error bars of the independent gas/dust determinations. This indicates that one can use these anisotropic Jeans models to recover the inclinations of the fast-rotators by enforcing the requirement $\beta_z>0.05$, and this explains the good agreement with the determinations from isotropic models we found in \citet{cappellari06}.

\subsection{Relation with the tensor virial theorem}

The Jeans equations are useful when one needs a predictions for the spatial variation of the kinematics in galaxies. However the stars in galaxies also satisfy more general global relations between kinematic observables, the most important of which are the virial equations \citep[sec.~4.3]{binney87}. These are a tensor generalization of the well known virial theorem $2K=-W$, which relates the total kinetic energy $K$ and the potential energy $W$ of an isolated stellar system in equilibrium.

The small dependence of the measured $M/L$ on the anisotropy (\refsec{sec:ml_comparison}) can be qualitatively understood as a consequence of the tensor virial equations. For an axisymmetric galaxy in equilibrium the total kinetic energy is proportional to the galaxy mass. The ratio of the kinetic energy $K_{zz}$ along the direction of the symmetry $z$-axis and the kinetic energy $K_{xx}$ along a direction orthogonal to it, is only a function of the galaxy shape (equation~[4-91] of \citealt{binney87}). The effect of increasing the anisotropy $\beta_z$ is simply that of redistributing the same kinetic energy $K_{xx}$ from the tangential to the radial direction. This reduces the velocity second moments along the galaxy projected major axis, while correspondingly increasing them along the minor axis. The projected luminosity-weighted second moments, from which the $M/L$ is measured, remain nearly unchanged when averaged over the full galaxy image.

The virial equations also provide a qualitative understanding of why the observations can constrain the inclination of an axisymmetric galaxy (\refsec{sec:inc_recovery}). In fact they state that the ratio $K_{xx}/K_{zz}$ increases with galaxy flattening. In other words, for a given projected surface brightness, a flat system needs more kinetic energy in a direction parallel to its the equatorial plane, to be in hydrostatic equilibrium.

The fast rotator NGC~2974 of \reffig{fig:models_beta_inc} shows high velocity second moments along the projected major axis. Although the galaxy does not appear very flat in projection, its kinematics is already an indication that the system is likely to be intrinsically flat. A flat model has more kinetic energy in the equatorial plane and will be able to easily reproduce the high $V_{\rm rms}$ along the major axis, without invoking extreme tangential anisotropy. A quantitative example, using both the first and second velocity moments, of how the tensor virial equations can be used to constrain the galaxy inclination is presented in fig.~11 of \citet{cappellari07}. There we showed that the \vse\ diagram of \citet{binney05} can be used to provide a qualitative estimate of the inclination of fast-rotators galaxies, given some assumptions on the anisotropy.

Although the virial equations can be used to estimate the $M/L$ and the inclination of galaxies, they have some obvious limitations: (i) They do not allow one to make full use of the spatial information contained in the integral-field observations. The whole two-dimensional kinematics is reduced to one number; (ii) They do not allow the limited spatial extent of the kinematics to be properly taken into account; (iii) They do not easily deal with non-similar isophotes and multiple photometric components in galaxies.

\section{Summary}
\label{sec:summary}

We present a generalization of the widely used semi-isotropic (two-integral) axisymmetric Jeans modeling method to describe the stellar dynamics of galaxies. Our method uses the powerful Multi-Gaussian Expansion (MGE) technique to accurately parameterize the galaxies photometry. It represents an anisotropic extension of what was presented in the semi-isotropic case by \citet{emsellem94}, and it maintains its simplicity and computational efficiency. We assume (i) a constant mass-to-light ratio and (ii) a velocity ellipsoid which is aligned with the cylindrical $(R,z)$ coordinates and has a flattening quantified by the classical $z$-anisotropy parameter $\beta_z=1-\overline{v_{z}^2}/\overline{v_{R}^2}$, where $z$ is the galaxy symmetry axis.

We test the technique using \sauron\ integral-field observations of the stellar kinematics \citep{emsellem04} for a small set of fast-rotator galaxies with a variety of kinematical properties. For galaxies that are constrained by the photometry to be close to edge-on we find that, although in the semi-isotropic limit ($\beta_z=0$) the models do not provide a good description of the data, the variation of the {\em single} global anisotropy $\beta_z$ is generally sufficient to accurately predict the shape of both the first ($V$) and second ($V_{\rm rms}=\sqrt{V^2+\sigma^2}$) velocity moments, once an detailed MGE parametrization of the photometry is given. An accurate description of the photometry, including ellipticity variations and disky isophotes, appears crucial to reproduce in detail the features of the kinematics.

In all cases we find that $\beta_z\ga0.05$, while generally $\sigma_{\phi}^2\approx\overline{v_{R}^2}$ with good accuracy. This confirms previous findings on the dynamical structure of these galaxies, showing that their velocity ellipsoid tends to be oblate \citep{cappellari07}. The anisotropy derived with our anisotropic Jeans dynamical modeling method agrees within the errors with the one previously measured using a more general Schwarzschild approach.

For fast-rotator galaxies that are {\em not} constrained by the photometry to be close to edge-on, we find that in general the inclination $i$ (or the corresponding galaxy shape) and the anisotropy $\beta_z$ are highly correlated and cannot be independently determined. However, if we introduce the observationally-motivated constraint $\beta_z\ga0.05$, the inclination becomes constrained to a narrow range of values and it agrees with independent determinations when those are available.

We are applying this method to determine the inclination, the mass-to-light ratio and the amount of rotation of a large sample of galaxies for which integral field kinematics are available. For galaxies close to edge-on, the global anisotropy or the dynamical structure of different galaxy subcomponents (e.g.\ bulge and disk) can also be investigated. We are using this method to test independent determinations of the masses of supermassive black holes in the nuclei of fast-rotator galaxies. This technique is ideal to study the dark matter content and the anisotropy of disks of spiral galaxies.

\section*{Acknowledgements}

I acknowledge fruitful discussions and feedback from Eric Emsellem, Davor Krajnovi\'c and Glenn van de Ven. I am grateful to Roger Davies, Tim de Zeeuw and Remco van den Bosch for constructive criticism on the draft.
I thank Nicholas Scott for providing the MGE model of three of the galaxies presented in \reffig{fig:models_velocity}, before publication. I am grateful to the anonymous referee for useful comments. I acknowledge support from a STFC Advanced Fellowship (PP/D005574/1). The \sauron\ observations were obtained at the William Herschel Telescope, operated by the Isaac Newton Group in the Spanish Observatorio del Roque de los Muchachos of the Instituto de Astrof\'{\i}sica de Canarias.

\appendix
\section{Seeing and aperture convolution}
\label{sec:psf}

The first and second velocity moments have to be convolved with the instrumental point-spread-function (PSF) and integrated over the aperture used in the observations, before making comparisons with the observed quantities.
The PSF effects are particularly important when studying the nuclear regions of galaxies (e.g.\ to estimate the mass of supermassive black holes). The PSF-convolved velocity moments are given by the general formulas (e.g.\ equations~[51]-[53] of \citealt{emsellem94})
\begin{equation}\label{eq:conv_surf}
\Sigma_{\rm obs} = \Sigma \otimes {\rm PSF}
\end{equation}
\begin{equation}\label{eq:conv_mu1}
[\overline{v_{\rm los}}]_{\rm obs} =
\frac{(\Sigma \overline{v_{\rm los}}) \otimes {\rm PSF}}{\Sigma_{\rm obs}}
\end{equation}
\begin{equation}\label{eq:conv_mu2}
[\overline{v_{\rm los}^2}]_{\rm obs} =
\frac{(\Sigma \overline{v_{\rm los}^2}) \otimes {\rm PSF}}{\Sigma_{\rm obs}},
\end{equation}
with $\otimes$ representing convolution and where the normalized PSF is generally described as the sum of $Q$ Gaussians
\begin{equation}\label{eq:psf}
\mathrm{PSF}(R) = \sum_{i=1}^Q
{\frac{G_i}{2\pi\sigma_i^2} \exp
\left(
    -\frac{R^2}{2\sigma_i^2}
\right)},
\end{equation}
with $\sum_{i=1}^Q G_i=1$. To efficiently evaluate numerically the above convolutions up to large radii, and even for very flat models, one can evaluate the models predictions $\Sigma\overline{v_{\rm los}^2}$ and $\Sigma\overline{v_{\rm los}}$ on a grid linear in the logarithm of the elliptical radius and in the eccentric anomaly. This is done by defining a logarithmically-spaced radial grid $R_j$ and then computing the moments at the coordinate positions $(x',y')=(R_j\cos\theta_k,  q' R_j \sin\theta_k)$, for linearly spaced $\theta_k$ values, with $q'$ a characteristic (e.g. the median) observed axial ratio of the MGE model. The model is then re interpolated onto a fine Cartesian grid before the convolution with the PSF using fast Fourier methods. Finally the model is rebinned into the observed apertures.

Alternatively, especially when the apertures sparsely sample the plane of the sky, instead of performing the convolutions of equations~\ref{eq:conv_surf}--\ref{eq:conv_mu2} on a regular grid, one can evaluate the same convolution integrals only for the observed apertures, while also including the integration onto the apertures \citep[appendix~D of][]{qian95}. Given a rectangular aperture of sides $L_x$ and $L_y$, aligned with the coordinate axes at position $(x',y')$ on the sky, and assuming a PSF given by \refeq{eq:psf}, the PSF-convolved observable $S(x',y')$ inside the aperture is
\begin{equation}\label{eq:convol}
S_{\rm obs}(x',y') = \int_{-\infty}^{\infty} \int_{-\infty}^{\infty}\!\!
    S(x',y')\, K(x-x',y-y')\, \dd x \dd y,
\end{equation}
where $S$ has to be replaced by $\Sigma$, $\Sigma\overline{v_{\rm los}}$, or $\Sigma\overline{v_{\rm los}^2}$ respectively, and correspondingly $S_{\rm obs}$ becomes $\Sigma_{\rm obs}$, $[\Sigma\overline{v_{\rm los}}]_{\rm obs}$, or $[\Sigma\overline{v_{\rm los}^2}]_{\rm obs}$ respectively. The kernel is given by
\begin{eqnarray}
K(x,y) & = & \sum_{i=1}^Q  \frac{G_i}{4}
\left\{\text{erf}\left[\frac{(L_x/2)-x}{\sqrt{2} \sigma_i }\right]+\text{erf}\left[\frac{(L_x/2)+x}{\sqrt{2} \sigma_i }\right]\right\}\nonumber\\
& \times & \left\{\text{erf}\left[\frac{(L_y/2)-y}{\sqrt{2} \sigma_i }\right]+\text{erf}\left[\frac{(L_y/2)+y}{\sqrt{2} \sigma_i }\right]\right\}.
\end{eqnarray}
In practice the integral of \refeq{eq:convol} can be limited to the region where $K(x,y)$ is significantly nonzero:
\begin{eqnarray}
-(L_x/2)-3\max\{\sigma_i\} \la x \la (L_x/2)+3\max\{\sigma_i\}\\ -(L_y/2)-3\max\{\sigma_i\} \la y \la (L_y/2)+3\max\{\sigma_i\}.
\end{eqnarray}

\end{document}